\documentclass[preprint,review,3p,times,sort,compress]{elsarticle}
\usepackage{etoolbox}

\usepackage{etoolbox}

\apptocmd{\thebibliography}{\setlength{\itemsep}{0.5em}}{}{}
\usepackage{amssymb}
\usepackage{amsmath}
\usepackage{array}	
\usepackage{booktabs}	
\usepackage{multirow}
\usepackage{multicol}
\usepackage{color}
\usepackage{xcolor}
\usepackage{colortbl}
\usepackage{hyperref}
\definecolor{mygray}{gray}{.9}
\definecolor{Revision}{RGB}{0, 0, 0}
\journal{Pattern Recognition}

\begin{document}

\begin{frontmatter}



\title{Scale-aware Adaptive Supervised Network with Limited Medical Annotations}


\author[1,2]{Zihan Li\fnref{equal}}
\fnmark[1]
\affiliation[1]{organization={Xiamen University},
                city={Xiamen},
                postcode={361005}, 
                country={China}}
\affiliation[2]{organization={University of Washington},
                city={Seattle},
                postcode={WA 98195}, 
                country={USA}}
\author[1]{Dandan Shan\fnref{equal}}

\author[3]{Yunxiang Li}
\affiliation[3]{organization={Department of Radiation Oncology, UT Southwestern Medical Center},city={Dallas},postcode={TX 75235}, 
                country={USA}}
                
\author[2]{Paul E. Kinahan}

\author[1]{Qingqi Hong\corref{cor1}}
\ead{hongqq@xmu.edu.cn}
\cortext[cor1]{Corresponding author}

\fntext[equal]{Zihan Li and Dandan Shan have equal contribution to this work.}

%% Abstract
\begin{abstract}
\textcolor{Revision}{Medical image segmentation faces critical challenges in semi-supervised learning scenarios due to severe annotation scarcity requiring expert radiological knowledge, significant inter-annotator variability across different viewpoints and expertise levels, and inadequate multi-scale feature integration for precise boundary delineation in complex anatomical structures. Existing semi-supervised methods demonstrate substantial performance degradation compared to fully supervised approaches, particularly in small target segmentation and boundary refinement tasks. To address these fundamental challenges, we propose SASNet (Scale-aware Adaptive Supervised Network), a dual-branch architecture that leverages both low-level and high-level feature representations through novel scale-aware adaptive reweight mechanisms. Our approach introduces three key methodological innovations, including the Scale-aware Adaptive Reweight strategy that dynamically weights pixel-wise predictions using temporal confidence accumulation, the View Variance Enhancement mechanism employing 3D Fourier domain transformations to simulate annotation variability, and segmentation-regression consistency learning through signed distance map algorithms for enhanced boundary precision. These innovations collectively address the core limitations of existing semi-supervised approaches by integrating spatial, temporal, and geometric consistency principles within a unified optimization framework. Comprehensive evaluation across LA, Pancreas-CT, and BraTS datasets demonstrates that SASNet achieves superior performance with limited labeled data, surpassing state-of-the-art semi-supervised methods while approaching fully supervised performance levels. The source code for SASNet is available at \href{https://github.com/HUANGLIZI/SASNet}{https://github.com/HUANGLIZI/SASNet}.}
\end{abstract}

\begin{graphicalabstract}
 \centering 
\includegraphics[width=\linewidth]{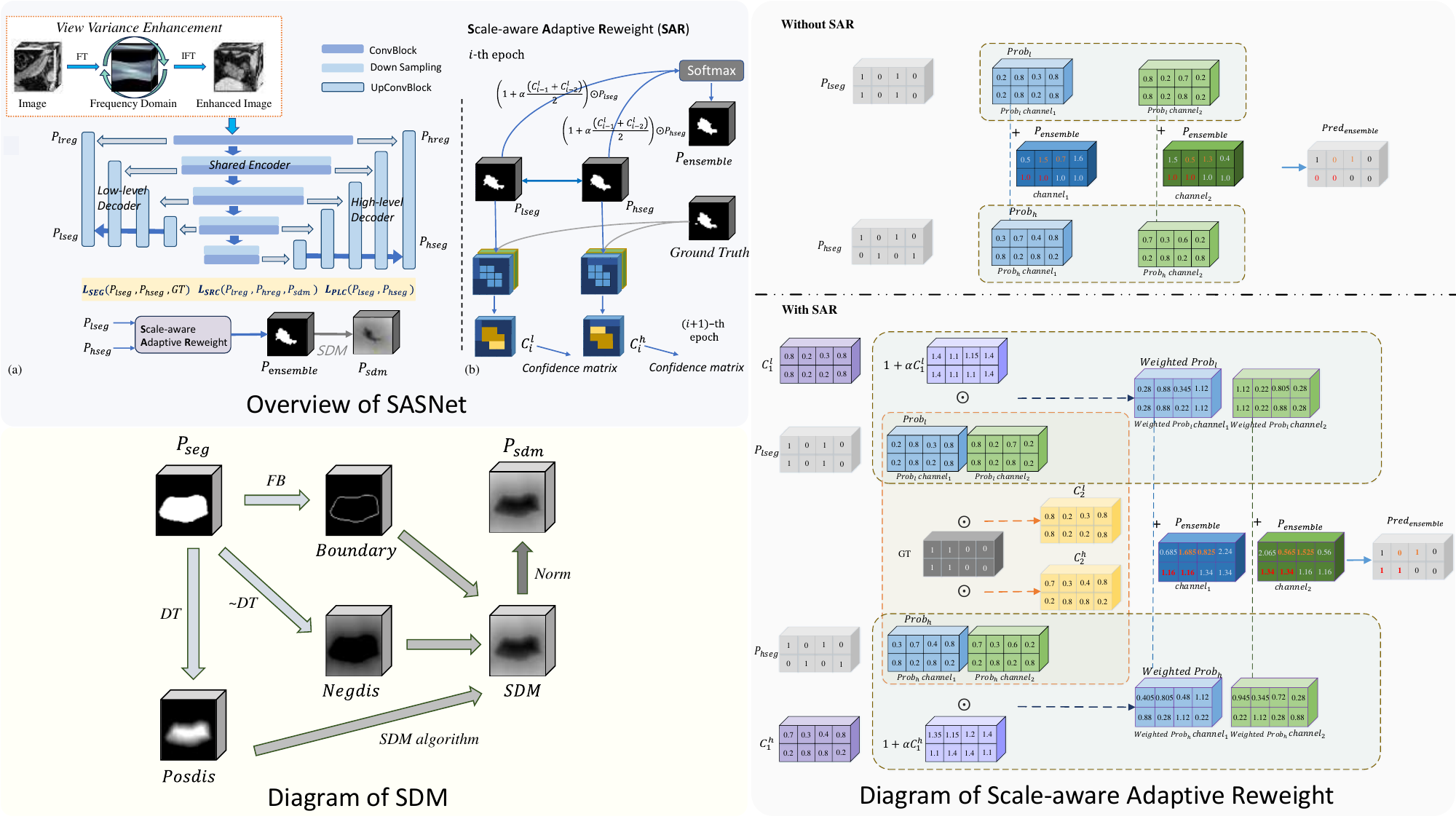}
  % \caption{Diagram of SDM (Signed Distance Map).}
\end{graphicalabstract}

\begin{highlights}
\item We propose SASNet, a dual-branch semi-supervised segmentation network that adaptively fuses multi-scale features to improve performance under limited annotation.
\item A scale-aware adaptive reweight strategy is introduced to generate more reliable ensemble predictions by selectively fusing pixel-wise results.
\item A view variance enhancement mechanism simulates annotation differences across views and scales, improving robustness and segmentation accuracy.
\end{highlights}

%% Keywords

\end{frontmatter}



\section{Introduction}

\label{sec:introduction}
In medical image segmentation, semi-supervised learning is crucial because high-quality dense annotations are both expensive and limited. At the same time, due to differences in the level of annotations, there may be some domain offset between them, resulting in less label information that can be used \cite{li2025visionunite}. Therefore, more and more researchers have begun to combine semi-supervised learning with medical image segmentation in recent years. Among them, Chaitanya et al. \cite{chaitanya2023local} designed a local contrast loss to help the model learn target features and generate better pseudo labels. In addition, many researchers have also begun to study the introduction of regularization terms into the loss function to help improve the performance of semi-supervised learning. Luo et al. \cite{luo2021efficient} constructed a joint framework that utilizes CNN and Transformer structure to learn different features of images and uses their prediction results for mutual supervision. You et al. \cite{you2022simcvd} used the teacher network and student network to calculate the comparative loss between the predicted results of the two networks. At the same time, researchers are also attempting to utilize multi-scale learning to improve the performance of medical image segmentation. Liu et al. \cite{liu2022region} directly input data of different scales into the encoder and output prediction results of different scales in the decoder section. Wang et al. \cite{wang2022multimodal} developed a multi-scale fusion module, which helps to fuse spatial information of different scales through route convolution. But researchers often ignore the role of multi-scale learning in semi-supervised learning \cite{tarvainen2017mean, zhou2019semi, li2023lvit}, so we design a new multi-scale learning paradigm\textemdash\textbf{Scale-aware Adaptive Learning} and innovatively introduce it into semi-supervised learning.
%\vspace{-1mm}
\begin{figure}[!t]
\centering
  \includegraphics[width=0.7\linewidth]{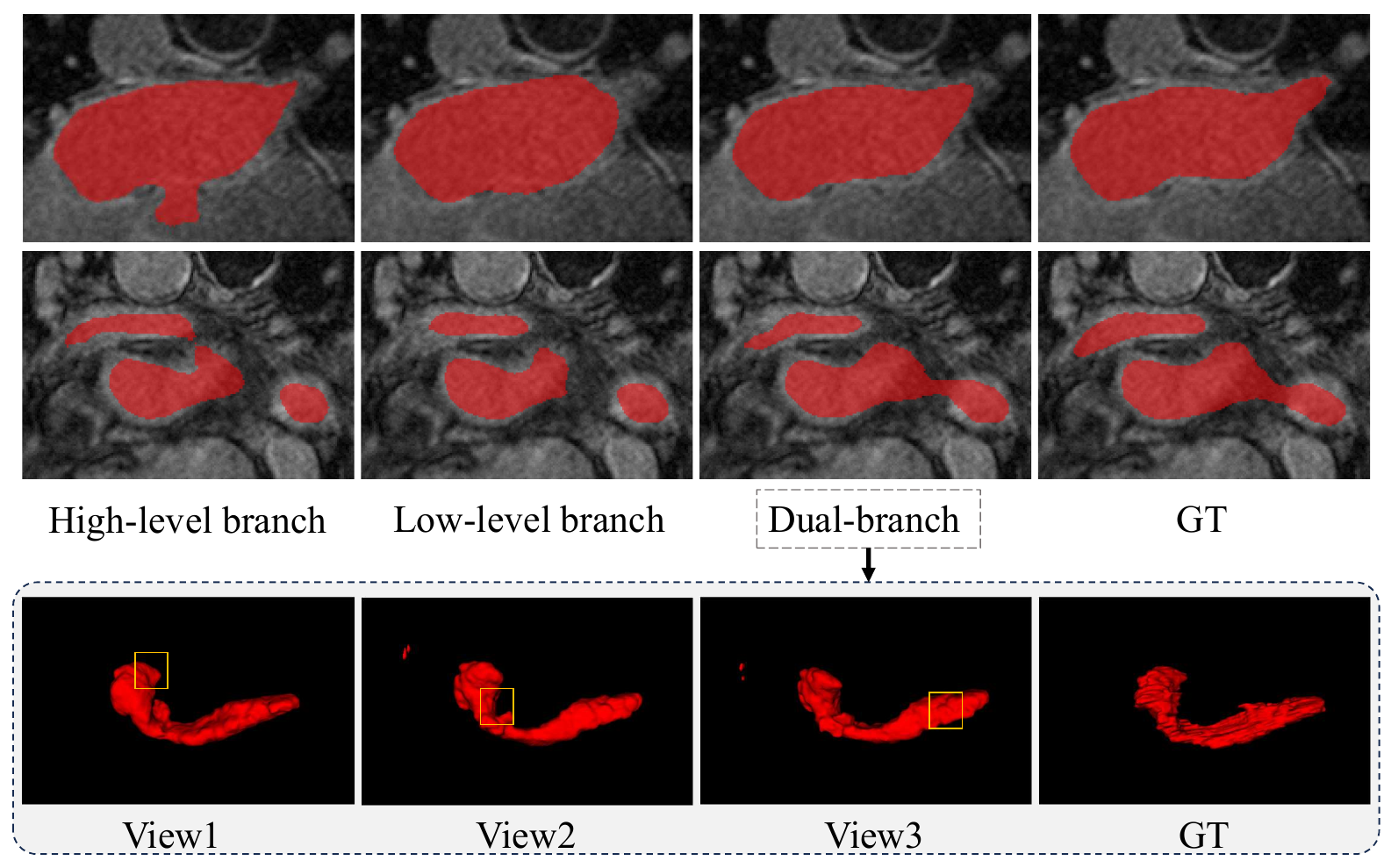}
  \caption{Comparison of segmentation results between different branch networks and multi-view inputs in the dual-branch network. The dual-branch network structure performs better in some details and is closer to the Ground Truth values than the low-level branch and high-level branch approach. Additionally, our model exhibits distinct prediction styles across different views, akin to the natural variability among annotations.
  %It benefits from limited labeled data, aiming to enhance the model's robustness by incorporating a form of annotator-style diversity.
  }
  \label{diff_branch_compare}
\end{figure}

As we mentioned above, there are still many unresolved issues in semi-supervised medical image segmentation. Firstly, \textbf{the challenge of missing labels} remains a significant issue, as semi-supervised methods continue to demonstrate lower performance in comparison to fully supervised approaches. Specifically, the segmentation of small targets and boundaries remains suboptimal, so we decide to further explore multi-scale information in the data to compensate for the missing labels. Incorporating feature information at multiple scales can enhance the model's ability to identify small targets and refine segmentation boundaries. Secondly, \textbf{the problem of annotation variance} poses a unique challenge in the field of medical imaging, given the varying focus. The difference in annotations can exacerbate the challenge of limited labeled data, especially when we need to refine predictions of boundaries. 

To address the above issues, we propose an 
Adaptive Supervised Hierarchical Network based on Scale Invariance.
%(SASNet). 
Specifically, to enhance the model's proficiency in learning multi-scale information, we design two different branches, one focusing on low-level features and the other on high-level features. Unlike previous multi-scale learning networks, we do not use multi-scale data input to different encoders but instead utilize multi-scale encoding features input to different decoders. We believe that it can obtain high-quality encoding features. In addition, directly utilizing encoding features at different scales can more intuitively utilize the information at different scales and display their common areas of interest as shown in Fig. \ref{diff_branch_compare}, which we call it \textbf{Scale Invariance}. The figure illustrates that both low-level and high-level features are utilized separately, ignoring the predicted connectivity information. Using them simultaneously can achieve the precise localization of the target, and enrich the model with more semantic information. As demonstrated in Fig. \ref{diff_branch_compare}, our model exhibits distinct prediction styles across different views, akin to the natural variability among predictions. It benefits from limited labeled data, aiming to enhance the model's robustness by incorporating view variance.
To address the above challenges, a natural consideration arises—integrating the results from both branches to achieve a more comprehensive and accurate segmentation. However, a simple addition of the two results may introduce undesirable noise, potentially compromising overall performance.
Therefore, we introduce the \textbf{Scale-aware Adaptive Reweight (SAR)} strategy. This approach involves pixel-wise weighting of predictions from both branches during the training process, based on the confidence acquired from preceding epochs. This adaptive mechanism empowers the network to selectively favor more reliable results, mitigating potential errors introduced through direct addition. The incorporation of the SAR strategy refines and controls the fusion process, resulting in a notable enhancement in overall segmentation performance.
To address the challenge of annotation variance, we introduce a view variance enhancement approach to simulate annotation differences. As shown in Fig. \ref{diff_branch_compare}, by changing the view of input, we enable the model to output segmentation results from different views of the same sample, approximating the annotation differences. Furthermore, we consider the segmentation results of branches at different scales as the annotation differences. This view variance enhancement strategy introduces diversity into the annotations, capturing the nuanced perspectives with varying viewpoints. By integrating complementary information from different scales and views, our approach comprehensively accounts for the inherent variations in annotations. 
Overall, our study proposes a novel semi-supervised learning approach that incorporates adaptive learning and cross-supervised learning and addresses the challenge of annotation variance via view variance enhancement, of which the main contributions are as follows:

\begin{itemize}

\item We propose a new segmentation network SASNet, which innovatively adopts a scale-aware adaptive reweight strategy to optimize pixel-wise results from different branches, generating more reliable ensemble predictions.

\item We propose an innovative view variance enhancement mechanism, synergistically merged with multi-scale branches, effectively emulating the annotation variance. This enhancement augments the resilience of semi-supervised learning, subsequently elevating the model's segmentation performance.

\item We evaluate our SASNet and other SOTA methods on three public datasets: the LA dataset, the Pancreas-CT dataset and the BraTS dataset. The results show that SASNet can outperform existing semi-supervised methods and achieve performance comparable to fully supervised methods.
\end{itemize}

\begin{figure*}[!ht]
  \centering
  \includegraphics[width=0.8\textwidth]{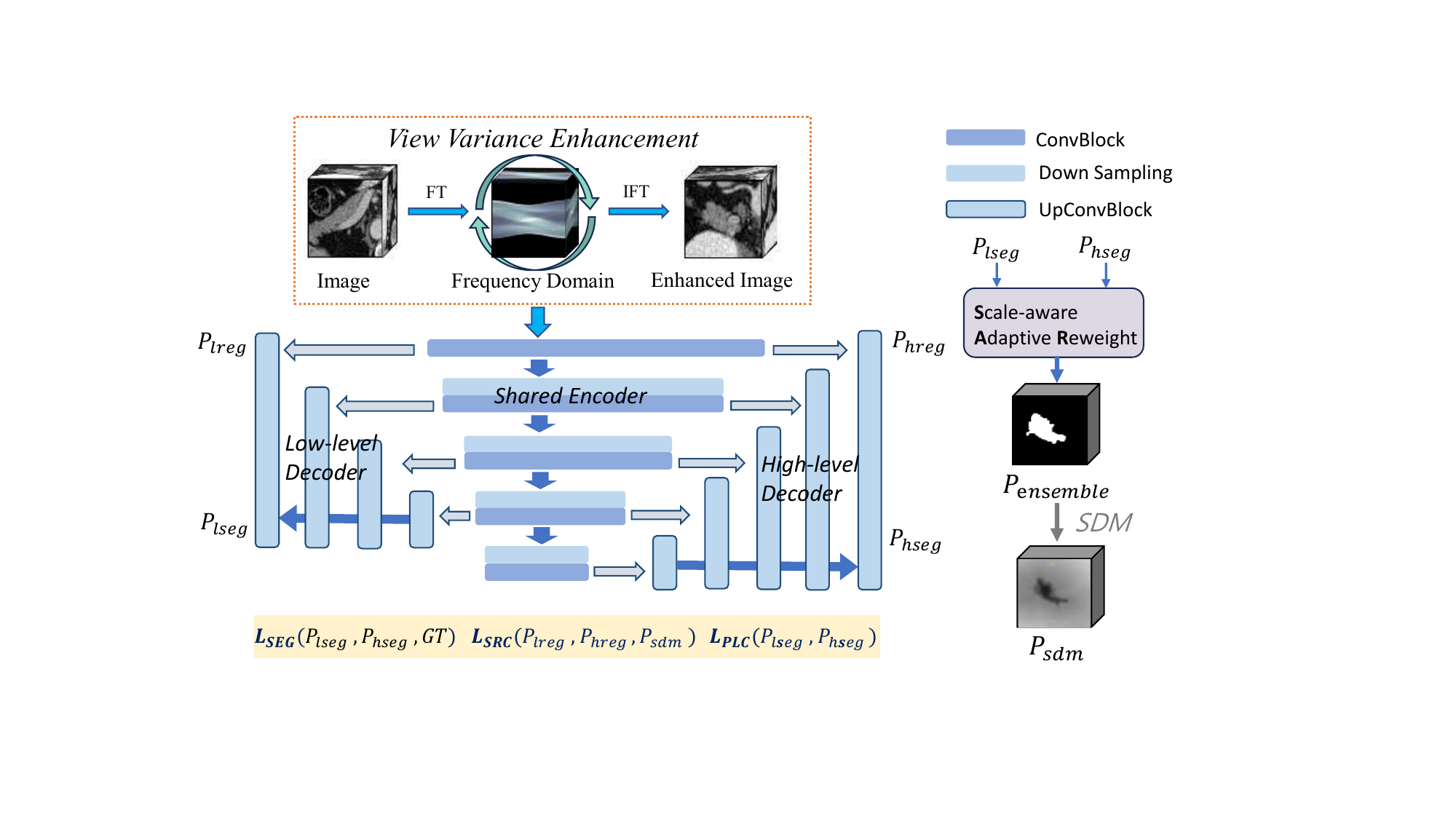}
  \caption{\textcolor{Revision}{Overview of SASNet. SASNet consists of three key components: the Dual-branch Architectural Network, the View Variance Enhancement Mechanism, and the Scale-Aware Adaptive Reweight Strategy. FT and IFT represent 3D Fourier Transform and 3D Inverse Fourier Transform, respectively. The training of SASNet is under the supervision of PLC Loss, SRC Loss, and SEG Loss. }}
  \label{SASNet}
  \vspace{-4mm}
\end{figure*}

\section{Related work}
\subsection{Semi-supervised Learning}
Traditional semi-supervised learning methods can be divided into self-training and consistent regularization \cite{zhang2025intra, qi2024gradient}. Self-training is a method to improve the performance of semi-supervised models by using a trained model with labeled data to predict pseudo-labels with high confidence on unlabeled data. Chaitanya et al. \cite{chaitanya2023local} designed a local contrast loss to learn features favorable for segmentation from the generated pseudo-labels during the self-training of the model. Adiga et al. \cite{adiga2024anatomically} proposed an anatomically-aware  framework that leverages unlabeled data for medical image segmentation. Ma et al. \cite{ma2024constructing} introduced a mixed-domain strategy that constructs intermediate domains to bridge the gap between labeled and unlabeled data from heterogeneous sources. Qi et al. \cite{qi2024gradient} developed a gradient-aware framework targeting class imbalance problems by adaptively modifying gradient updates based on class frequencies.
\textcolor{Revision}{Recent advances in semi-supervised medical image segmentation have explored the collaboration mechanisms and versatile paradigms for leveraging unlabeled data. Zeng et al. \cite{zeng2024reciprocal} introduced reciprocal collaboration frameworks that enable bidirectional knowledge exchange between network components, while their subsequent work \cite{zeng2025segment} proposed versatile paradigms that adaptively segment diverse anatomical structures through unified architectural designs. The PICK framework \cite{zeng2025pick} employs prediction-masking strategies to selectively utilize confident pseudo-labels during training. While these approaches demonstrate promising results through collaborative learning and adaptive masking, they fundamentally differ from our methodology in several critical aspects. Unlike reciprocal collaboration mechanisms that rely on symmetric knowledge exchange, our Scale-aware Adaptive Reweight (SAR) strategy introduces asymmetric, confidence-driven weighting that dynamically adjusts pixel-wise predictions using historical performance matrices from preceding epochs. Furthermore, our approach uniquely integrates 3D Fourier domain transformations for view variance enhancement, simulating annotation variability through frequency-domain manipulations rather than conventional data augmentation.}

\subsection{Consistency Learning}
In the framework of consistency regularization, existing methods enhance the generalization ability of models by enforcing consistency in model predictions under different data augmentations \cite{shan2025stpnet}. Yu et al. \cite{yu2019uncertainty} proposed an uncertainty-guided Mean Teacher framework, which combines transformation consistency to improve performance. Luo et al. \cite{luo2021efficient} introduced a pyramid consistency regularization framework with uncertainty correction, extending the basic segmentation network to generate pyramid predictions at different scales, and supervising unlabeled images with multi-scale consistency loss to ensure consistency in predictions at different scales for the same input. Bai et al. \cite{bai2023bidirectional} proposed a bidirectional copy-paste method aimed at combining labeled and unlabeled data, implemented within a simple Mean Teacher framework. This method effectively reduces the gap between empirical distributions by encouraging unlabeled data to learn integrated common semantics from labeled data.
\textcolor{Revision}{There are significant advances in uncertainty-driven consistency mechanisms and multi-constraint optimization frameworks. The uncertainty-participation context consistency learning approach \cite{yin2025uncertainty} introduces sophisticated uncertainty quantification methods that adaptively weight consistency losses based on prediction confidence, demonstrating how uncertainty estimation can enhance pseudo-label reliability. Similarly, recent developments in stereo vision applications have explored decoupling-coupling paradigms \cite{wei2025decoupling} that separate certainty estimation from primary task learning while maintaining architectural cohesion through carefully designed coupling mechanisms. Multi-constraint consistency learning frameworks \cite{yin2025semi} further advance this paradigm by incorporating diverse consistency objectives that operate across multiple semantic and spatial scales, creating robust optimization landscapes for semi-supervised training. These advances establish important precedents for uncertainty-aware learning and multi-objective consistency optimization. However, our proposed Scale-aware Adaptive Reweight (SAR) strategy introduces distinct innovations that extend beyond conventional uncertainty-participation frameworks. Unlike static uncertainty weighting schemes, SAR employs temporal confidence accumulation across training epochs, creating dynamic pixel-wise weighting matrices that capture long-term prediction reliability patterns rather than instantaneous uncertainty estimates.}

\subsection{Multi-scale Medical Image Segmentation}
Due to abundant noise and intricate morphology in medical images, combining multi-scale information encourages models to capture lesion features across various scales \cite{li2022tfcns}. Existing multi-scale medical image segmentation methods can be divided into multi-scale methods between different layers and multi-scale methods in the same layer \cite{shan2023coarse}. For the multi-scale methods between different layers, the contextual information between different layers is usually learned using \cite{li2024ai}. Luo et al. \cite{luo2021efficient} constructed a pyramid prediction network to learn unlabeled data by minimizing the difference between segmented predictions at different scales and their mean values. Liu et al. \cite{liu2022region} input samples of different scales at each layer of the encoder and output features of different scales in the decoder part for depth supervision. For multi-scale methods on the same layer, usually dilated convolution or pyramid pooling is used to learn multi-scale features of images on the same layer. Wang et al. \cite{wang2022multimodal} inserted a multiscale background fusion module into UNet to fuse the spatial information at different scales. However, these methods mentioned above only perform multi-scale fusion at the encoder. In our work, we utilize adaptive learning to perform scale-aware adaptive reweight ensemble for predictions at different scales.

\begin{figure*}[!ht]
  \centering
  \includegraphics[width=0.45\textwidth]{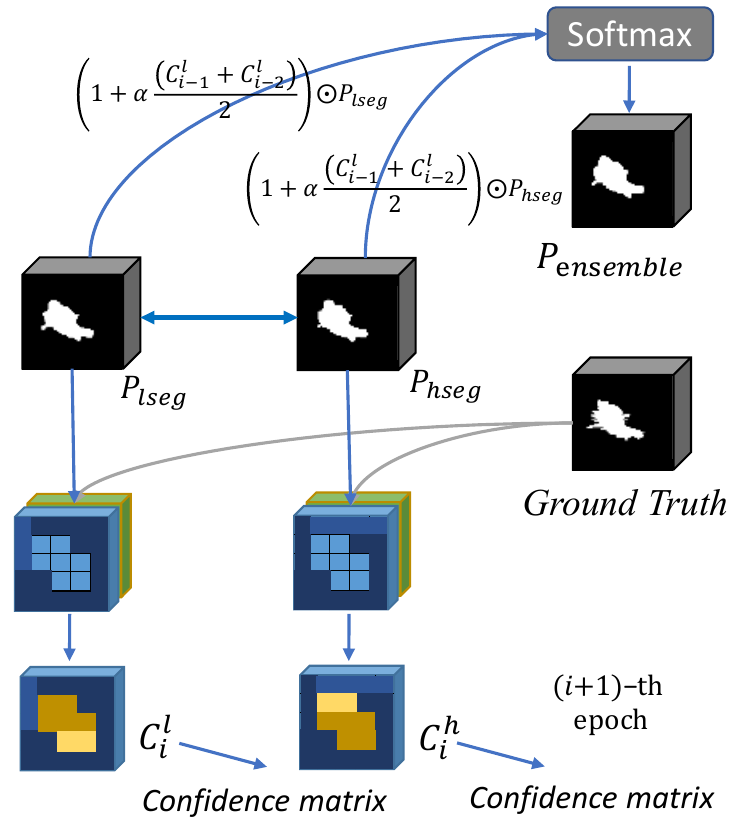}
  \caption{\textcolor{Revision}{Details of the Scale-Aware Adaptive Reweight (SAR) strategy at the i-th epoch.}}
  \label{SAR_Frame}
  \vspace{-4mm}
\end{figure*}

\section{Method}
The proposed SASNet, as illustrated in Fig. \ref{SASNet}, consists of three key components: the Dual-branch Architectural Network, Scale-Aware Adaptive Re-weighting Strategy, and the View Variance Enhancement Mechanism. For the semi-supervised training objectives of SASNet, we have formulated three types of objectives: traditional segmentation objectives using limited label information, consistency learning objectives between the individual regression prediction and the ensemble regression prediction using the SDM algorithm, and pseudo label cross-supervised learning objectives using different scale prediction.
\textcolor{Revision}{To implement these objectives, the training pipeline of SASNet proceeds as follows. The input 3D medical images are first processed by the View Variance Enhancement Mechanism to simulate annotation variability, and then fed directly into  the Dual-branch Architectural Network, where high-level and low-level features generate corresponding segmentation and regression outputs. The segmentation results from both branches are integrated via the Scale-Aware Adaptive Re-weighting Strategy, combined with temporal confidence accumulation to produce dynamically weighted pixel-level predictions. These weighted predictions, after being processed by the SDM algorithm, are further used to supervise the regression branch, enhancing boundary precision and improving predictive capability under semi-supervised conditions.}

\begin{figure*}[!h]
 \centering
  \includegraphics[width=0.9\linewidth]{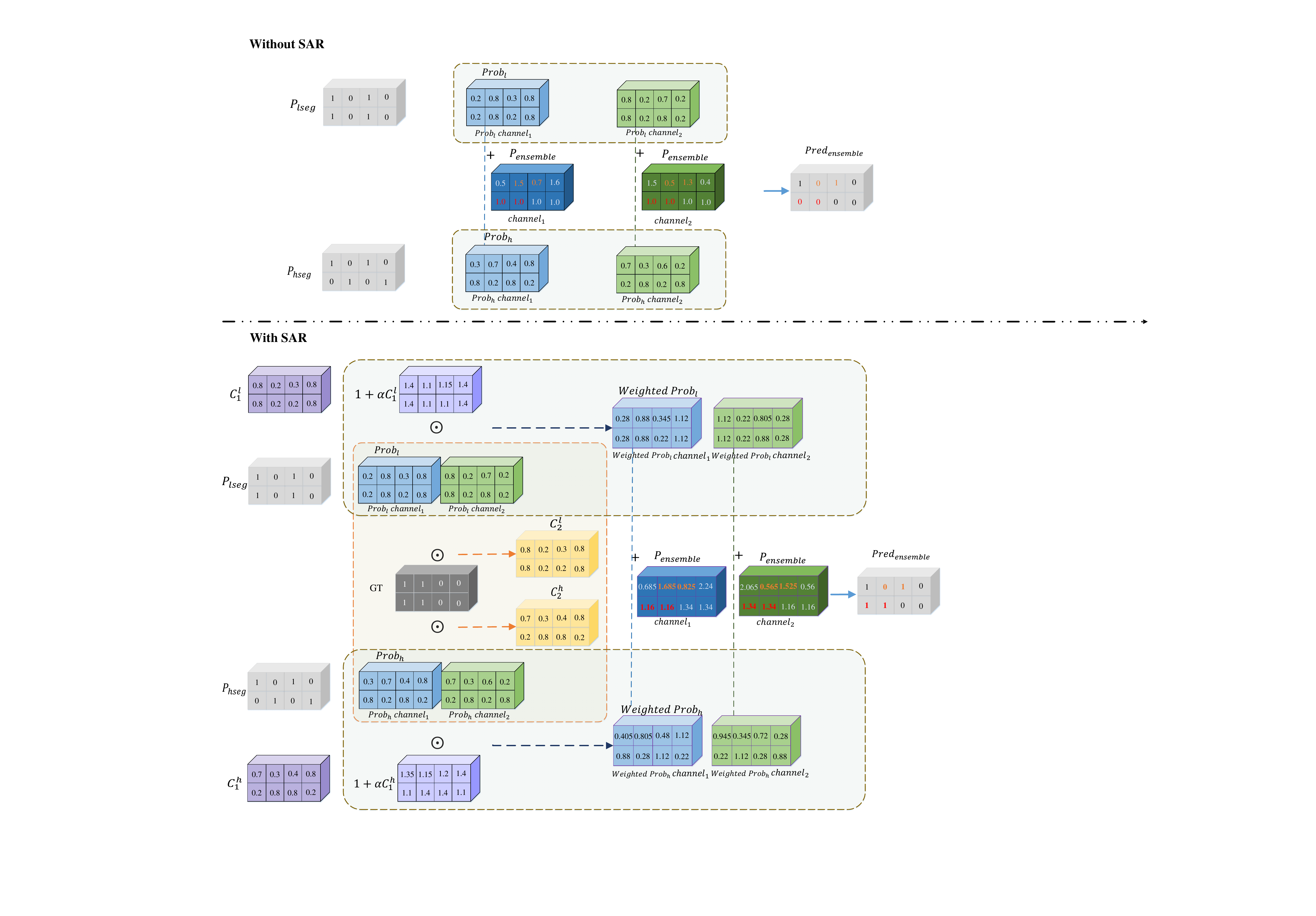}
  \vspace{-6mm}
  \caption{Diagram of Scale-aware Adaptive Reweight (SAR). Comparison of output results with and without SAR. Assuming that the predicted probabilities $Prob_{l}$ from the low-level branch and $Prob_{h}$ from the high-level branch remain unchanged. $C_{1}^{l}$ stands for the confidence matrix of the low-level branch in the first epoch. $\alpha$ represents the residual coefficient. Weighted $Prob_{l}$ and Weighted $Prob_{h}$ represent the probabilities after weighting for the low-level branch and high-level branch, respectively. $C_{ch1}^{l}$ denotes the first channel of $C_{1}^{l}$, while $C_{ch2}^{l}$ signifies the second channel. Similarly, $C_{ch1}^{h}$ and $C_{ch2}^{h}$ stand for the first and second channels of $C_{1}^{h}$, respectively.}
  \vspace{-4mm}
  \label{SAR}

\end{figure*}

\subsection{Dual-branch Hierarchical Network}
To address the challenges of multi-scale and semi-supervised learning, we design a Dual-branch Hierarchical Network, which comprises a shared encoder and two decoders for decoding low-level and high-level features, respectively. The low-level decoder is responsible for decoding small-scale features, and the high-level decoder is responsible for decoding large-scale features. Prior to inputting the data into the network, view variance enhancement is employed to create multiple view volumes, from which we randomly select one for processing. The shared encoder is constructed using ConvBlock and downsampling layers that are interleaved. The input image is first convolved to generate image features, which are then downsampled. The output of downsampling serves as input to the next convolutional layer. The features from layers 1 to 4 are passed to the low-level decoder, and the features from layers 1 to 5 are passed to the high-level decoder. The outputs of the fourth layer and the final layer are passed to the low-level and high-level decoders, respectively, for feature decoding. The UpConvBlock that constitutes the decoder includes a convolutional layer for decoding and a deconvolution layer for upsampling. Additionally, we retain residual connections to improve feature flow between the encoder and decoder.
\setlength{\abovedisplayskip}{0pt} % Space above the formula
\begin{eqnarray}
    &x_{i}^{'} = ConvBlock_{i-1}(x_{i-1})
    \label{3-1}\\
    &x_{i} = DownSampling_{i-1}(x_{i}^{'})
    \label{3-2}\\
    & y_{i-1} = UpConvBlock_{i-1}(y_{i})+x_{i-1}
    \label{3-3}
\end{eqnarray}
where $x_{i}^{'}$, $x_{i}$ denote the encoding feature after the $(i-1) $-th convolution layer and the output feature of the $(i-1) $-th downsampling layer respectively. And $y_{i-1}$ denotes the decoding feature of the $(i-1) $-th UpConvBlock, whose inputs are $y_{i}$ and $x_{i-1}$ respectively.

%\vspace{-2mm}
\subsection{Scale-aware Adaptive Reweight}
As shown in Fig. \ref{SAR_Frame}, we employ a Scale-aware Adaptive Reweight (SAR) strategy during the training process to generate more reliable ensemble prediction. Specifically, within each training epoch, for the labeled data, we compute the average confidence scores from the previous two epochs ($(i-1) $-th and $(i-2) $-th epoch) of low-level branch and high-level branch respectively. After applying a residual calculation, we individually weight the current predicted probabilities of these two branches. After summing the weighted probabilities, we input them into a softmax function to obtain a ensemble prediction, denoted as $P_{ensemble}$. By pixel-wise weighting of the predictive outcomes from two branches and achieving intelligent weight adjustment, our model is capable of adaptively learning features across various scales. Simultaneously, we multiply the current predicted probabilities of the two branches by the Ground Truth (GT) to derive the confidence score, $C_{i}$, for the $i$-th epoch. The confidence score $C_{i}$ for the $i$-th epoch can be employed in the subsequent round of calculations, aiming to further enhance the optimization of the model training process. As for the unlabeled data, we average the current predicted probabilities from the two branches, $P_{lseg}$ and $P_{hseg}$, to form the pseudo-label for unlabeled data. The reweighting strategy for labeled data can be represented as the following formulas:
%\vspace{-2mm}
\begin{eqnarray}
    &P_{lseg}^{'}=(1+ \alpha \frac{\left ( C_{i-1}^{l}+C_{i-2}^{l}\right )}{2})\odot P_{lseg}\\
    &P_{hseg}^{'}=(1+ \alpha \frac{\left ( C_{i-1}^{l}+C_{i-2}^{l}\right )}{2})\odot P_{hseg}\label{4-2}\\
    &P_{ensemble}=softmax(P_{lseg}^{'}+P_{hseg}^{'})
    \label{4-3}
\end{eqnarray}
$P_{lseg}$ represents the output predicted by the low-level decoder, while $P_{hseg}$ represents the output predicted by the high-level decoder. $C_{i-1}^{l}$ and $C_{i-2}^{l}$ respectively denote the confidence matrices of the low-level decoder for the $(i-1) $-th and $(i-2) $-th epochs, and $C_{i-1}^{h}$ and $C_{i-2}^{h}$ respectively denote the confidence matrices of the high-level decoder for the $(i-1) $-th and $(i-2) $-th epochs. $\alpha$ represents a residual coefficient within the range of 0 to 1. And the above strategy applies a weight of 0.5 as $\alpha$ to each branch empirically.

The computation of the confidence maps $C_{i}$ for the $i$-th epoch can be expressed with the following formulas:
\begin{eqnarray}
    &C_{i}^{l}=P_{lseg}^{0}\odot [GT==0]+P_{lseg}^{1}\odot [GT==1]\label{4-4}\\
    &C_{i}^{h}=P_{hseg}^{0}\odot [GT==0]+P_{hseg}^{1}\odot [GT==1]\label{4-5}
\end{eqnarray}
where $P_{lseg}^{0}$ and $P_{lseg}^{1}$ represent the values of the 0th and 1th channels of the output predicted by the low-level decoder, and $P_{hseg}^{0}$ and $P_{hseg}^{1}$ represent the values of the 0th and 1th channels of the output predicted by the high-level decoder. $C_{i}^{l}$ signifies the confidence map of the low-level decoder obtained in the $i$-th epoch, and $C_{i}^{h}$ represents the confidence map of the high-level decoder obtained in the $i$-th epoch.

Fig. \ref{SAR} illustrates a comparison between the scenarios with and without the use of SAR. It is evident from the graph that employing the SAR strategy can lead the predicted probabilities to converge towards the GT. Specifically, after the application of adaptive reweighting, the values corresponding to the first and second positions (highlighted in red) in the second row of $Pred_{ensemble}$ have transitioned from 0 to 1, aligning precisely with the GT. Following the implementation of SAR, although the second and third positions (highlighted in yellow) of the first row of $Pred_{ensemble}$ still do not match the GT, there has been a noticeable change in the values of channel1/channel2 at the second position. These values have decreased from 1.5/0.5 to 1.685/0.565. This indicates that $P_{ensemble}$ tends to increase the predicted probability for channel2 (foreground), leading the prediction towards 1. Similarly, a similar logic applies to the third position in the first row, where the values of channel1/channel2 have increased from 0.7/1.3 to 0.825/1.525. This suggests that $P_{ensemble}$ tends to increase the predicted probability for channel one (background), steering the expected result to 0.

\subsection{View Variance Enhancement Mechanism}
To simulate annotation differences and enhance the robustness of the model, we have devised a mechanism for enhancing view variance. Initially, we employ 3D Fourier Transform (FT) to map the input 3D images into the frequency domain. Subsequently, a series of data augmentation operations, such as rotation, are performed in the frequency domain to generate a sequence of images with different viewpoints. These operations enable the model to encounter a richer variety of viewpoint changes during the training phase, thereby improving its robustness to input images from different viewpoints. Then, we use the Inverse Fourier Transform (IFT) to map these transformed images back to the time domain. In our approach, we sequentially select three viewpoints as inputs. We hypothesize that the prediction results from different viewpoints reflect the differences in annotations under the same viewpoint. Therefore, exposing the model to such differences during training can enhance its robustness and adaptability to complex annotations and diverse inputs.

This method not only simulates annotation differences in the real world but also enhances the model's generalization ability to unknown viewpoint inputs, making it more reliable and effective in practical applications.
\textcolor{Revision}{Different from conventional spatial-domain augmentation or standard consistency regularization methods, this mechanism is capable of generating image transformations that reflect clinical annotation differences, making the training process more specifically targeted at the inherent uncertainty in real annotations, rather than merely increasing data diversity. Specifically, operations in the frequency domain can simultaneously affect both local and global image features, enabling comprehensive and coherent transformations while preserving critical anatomical structures and naturally maintaining the fundamental harmonic relationships in medical images, thereby avoiding interpolation artifacts commonly introduced by spatial rotations. In addition, this mechanism, combined with the dual-branch architecture, generates controlled prediction variations, using these variations as proxies for inter-annotator disagreement, which allows the model to learn robust features despite annotation uncertainty.}

The specific operations can be represented by following formulas:
\begin{eqnarray}
    & I_f = \iiint I \cdot e^{-2\pi i (k_x x + k_y y + k_z z)} dx dy dz \\
    & I_{f'} = \text{rotate}(I_f, \theta) \\
    & I' =  \iiint I_{f'} \cdot e^{2\pi i (k_x x + k_y y + k_z z)} dk_x dk_y dk_z
\end{eqnarray}
where $I$ represents an input 3D image, with $x, y, z$ representing the three coordinates in space, and $k_{x}, k_{y}, k_{z}$ representing the coordinates in the frequency domain. $I_{f}$ is the image mapped to the frequency domain after undergoing 3D Fourier Transform, and $I_f^{'}$ is the image obtained from rotating in the frequency domain, with $\theta$ representing the rotation angle.

\begin{figure}[!t]
 \centering 
  \includegraphics[width=0.85\linewidth]{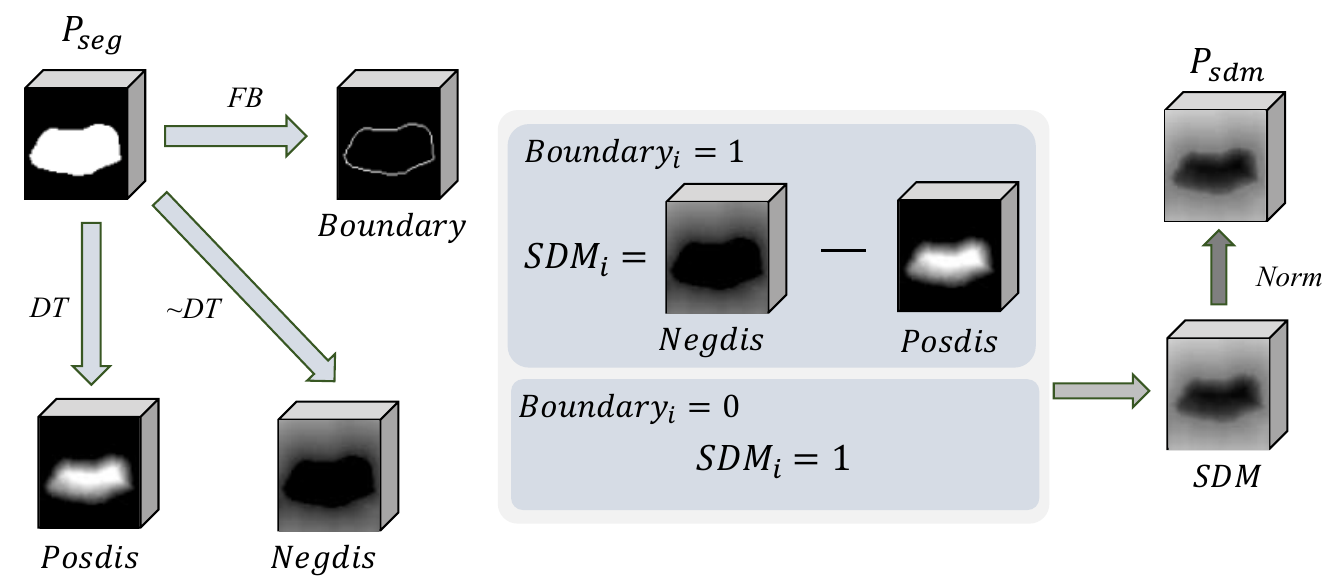}
\caption{\textcolor{Revision}{Diagram of SDM (Signed Distance Map) Generation Pipeline. The workflow includes boundary extraction via feedback (FB), parallel distance transforms (DT and  $\sim$DT) producing positive and negative distance maps.}}
  \label{SDM}
\end{figure}

\subsection{Semi-supervised Training Objectives}
To achieve effective semi-supervised training of SASNet, we have designed multiple training objectives to supervise the training process. The first is Limited Annotation Learning. We use limited label information $GT$ to monitor the training of dual branches through $L_{seg}$. Followed by Pseudo Label Cross-supervised Learning the segmentation predictions of the two branches are supervised through $L_{plc}$ as pseudo-label of the other segmentation prediction, respectively. Finally, Segmentation-Regression Consistency Learning is used to use integrated prediction $P_{ensemble}$ and SDM algorithms to form $P_{sdm}$ to supervise the regression predictions of the two branches through $L_{src}$. \textcolor{Revision}{The total loss function is as follows: ${L}_{Semi}=\beta*{L}_{seg}+\gamma*({L}_{plc}+{L}_{src})$, where the weight coefficient $\beta$ is 0.5. And during the training process of SASNet, the consistency coefficient $\gamma$ is increased by the sigmoid ramp-up function \cite{laine2017temporal} as the training epochs increase. The final consistency coefficient $\gamma$ is 1 after 40 training epochs.}

\subsubsection{Limited Annotation Learning}
In the supervision part, we utilize limited label information for supervision. As shown in Fig. \ref{SASNet},  GT and $L_{seg}$  supervise the prediction output of the low-level branch and the high-level branch, respectively. The SEG loss we utilize is Dice loss and cross-entropy loss, as shown below:
\begin{eqnarray}
    &{L}_{seg}=L_{Dice}+L_{CE}
    \label{3.3.1}\\
    &L_{Dice}=1-\sum_{i=1}^{N}{\frac{1}{N}\cdot\frac{2\left|p_{i}\cap y_{i}\right|}{\left(\left|p_{i}\right|+\left|y_{i}\right|\right)}}\\
    &L_{CE}=-\sum_{i=1}^{N}{\frac{1}{N}\cdot}y_{i}\log{\left(p_{i}\right)}
\end{eqnarray}
 where $p_{i}$ and $y_{i}$ represent the prediction and GT for the $i$-th voxel. The $N$ denotes the total number of voxels.

\subsubsection{Pseudo Label Cross-supervised Learning}

Different from the previous pseudo-label learning, the initial model is trained with labeled data, and then the prediction results of the initial model are used as the pseudo-label for unlabeled data. We use the prediction results of the two branches as the pseudo-labels of each other for cross-supervision and learning. The PLC Loss is calculated by the following formula:
\begin{eqnarray}
    &{L}_{plc}=\sum_{i=1}^{N}{\frac{1}{N}\cdot}(p_{lseg}-p_{hseg})^{2}
\end{eqnarray}
where $p_{lseg}$ and $y_{hseg}$ represent the segmentation prediction of $i$-th voxel by the low-level branch and high-level branch.

\subsubsection{Segmentation-Regression Consistency Learning}
We also introduce consistency learning for the training of SASNet. We first form the integrated segmentation prediction $p_{ensemble}$ through the SAR strategy. Then $p_{ensemble}$ is processed through SDM to form $p_{sdm}$. And $p_{sdm}$ is applied to the regression predictions $p_{lreg}$ and $p_{hreg}$ for segmentation-regression consistency learning. The loss function ${L}_{SRC}$ is delineated below:

\begin{eqnarray}
    &{L}_{src}=\sum_{i=1}^{N}{\frac{[(P_{sdm}-P_{lreg})^{2}+(P_{sdm}-P_{hreg})^{2}]}{N}}
\end{eqnarray}
where $p_{lreg}$ and $y_{hreg}$ represent the regression prediction of the $i$-th voxel by the low-level branch and the high-level branch respectively. $p_{sdm}$ denotes the output of $p_{ensemble}$ after the SDM processing.

The SDM procedure is illustrated in Fig. \ref{SDM}, ${Posdis}$ is formed by the distance$\_$transform$\_$edt $(DT)$ function from the segmentation prediction $P_{seg}$, and the result of anti-DT ($\sim$${DT}$) is ${Negdis}$. At the same time, the $Border$ is obtained through the Find$\_$boundaries $(FB)$ function. And then the $SDM$ is obtained by ${Posdis}$, ${Negdis}$, and $Border$, and finally forms $P_{SDM}$ through a normalized operation.

\section{Experiments}
\label{sec:experimental}
\subsection{Datasets}
\textbf{LA} \cite{xiong2021global}:
The left atrial dataset comprises 100 gadolinium-enhanced MR imaging scans with a uniform resolution of $0.625\times0.625\times0.625 mm^3$. To ensure consistency with prior work \cite{luo2021semi}, we use the same 80 of these images for training and 20 for validation. Prior to training, we preprocess the images by expanding the bounds to randomly selected values between 10-20, 10-20, and 5-10, respectively.

\textbf{Pancreas-CT} \cite{clark2013cancer}:
The pancreas dataset is collected by the National Institutes of Health Clinical Center (NIH) and contains 82 3D abdominal CT scans annotated by experts. We randomly select 62 of them for training and the rest 20 for validation. We first resample the spacing of the data to $1.0\times1.0\times1.0 mm^3$. Then, we crop the voxels with Hounsfield Units value from -125 to 275, expand the edges by 25, 25 and 0 voxels, respectively.

\textbf{BraTS} \cite{hdtd-5j88-20}:
The Brain Tumor Segmentation 2019 dataset comprises preoperative MRI scans from 335 glioma patients across multiple institutions. Each patient's MRI includes four modalities: T1, T1Gd, T2, and T2-FLAIR. We specifically utilize the T2-FLAIR modality for whole tumor segmentation due to its superior ability to highlight malignant tumors \cite{zeineldin2020deepseg}. All images are resampled to an isotropic resolution of $1.0\times1.0\times1.0 mm^3$.
For our experiments, we use 250 samples for training, 25 for validation, and the remaining 60 for testing.

\begin{figure*}[!ht]
  \setlength{\abovecaptionskip}{1mm}
  \includegraphics[width=\textwidth]{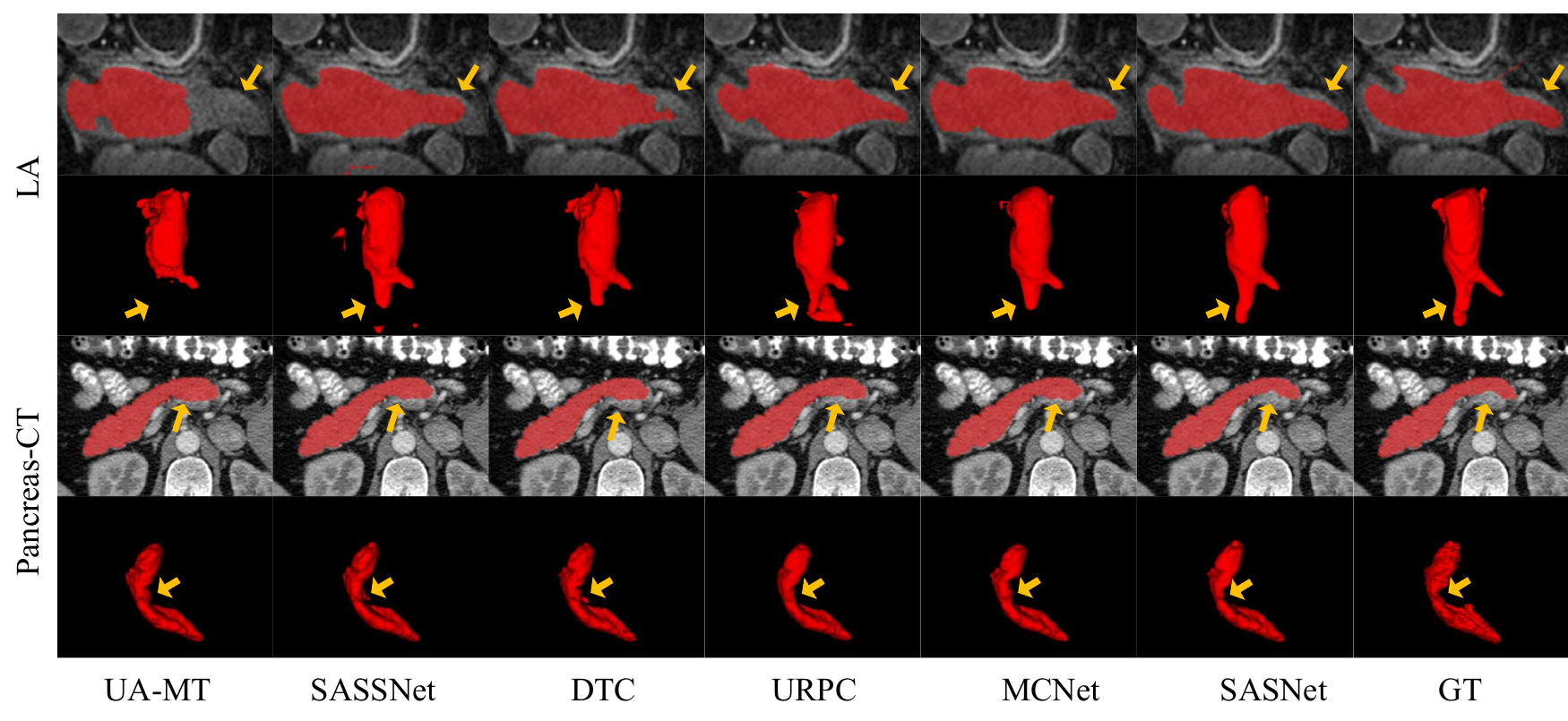}
  \caption{Qualitative results of SASNet and other methods on the LA dataset and Pancreas-CT dataset with 20\% of labeled data.}
  \label{LA_Pancreas_result}
\end{figure*}

\begin{figure*}[!t]
  \setlength{\abovecaptionskip}{2mm}
  \includegraphics[width=\textwidth]{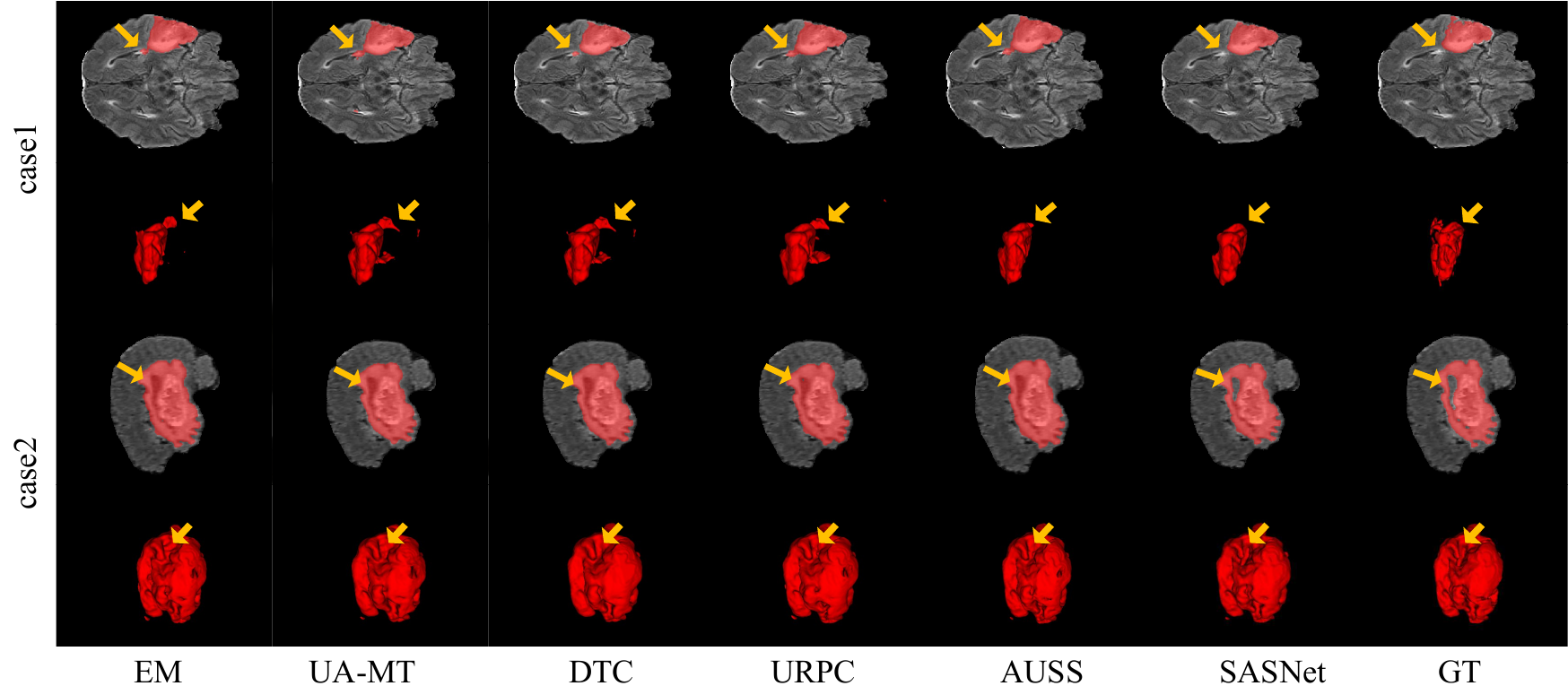}
  \caption{\textcolor{Revision}{Qualitative results of SASNet and other methods on the BraTS dataset with 20\% of labeled data.}}
  \label{BraTs_result}
\end{figure*}

\begin{table*}[!ht]
\footnotesize
  \centering
  \resizebox{0.85\columnwidth}{!}{%
    \begin{tabular}{crrcccc}
    \toprule
    \multirow{2}[0]{*}{Method} & \multicolumn{2}{c}{\#Scans used} & \multicolumn{4}{c}{Metrics} \\
    \cmidrule(r){2-3} \cmidrule(l){4-7}
          & \multicolumn{1}{c}{Labeled} & \multicolumn{1}{c}{Un} & Dice (\%) $\uparrow$ & Jaccard (\%) $\uparrow$ & HD95 (voxel) $\downarrow$ & ASD (voxel) $\downarrow$ \\
    \midrule
    V-Net \cite{milletari2016v} \textcolor{gray}{(3DV 2016)}& \multicolumn{1}{c}{8(10\%)} & \multicolumn{1}{c}{0} & 79.99 & 68.12 & 21.11 & 5.48 \\
    V-Net \cite{milletari2016v} \textcolor{gray}{(3DV 2016)}& \multicolumn{1}{c}{16(20\%)} & \multicolumn{1}{c}{0} & 86.03 & 76.06 & 14.26 & 3.51 \\
    V-Net \cite{milletari2016v} \textcolor{gray}{(3DV 2016)}& \multicolumn{1}{c}{80(100\%)} & \multicolumn{1}{c}{0} & 91.14 & 83.82 & 5.75  & 1.52 \\ \hline
    UA-MT \cite{yu2019uncertainty}  \textcolor{gray}{(MICCAI 2019)} &\multicolumn{1}{c}{8(10\%)} & \multicolumn{1}{c}{72}  & 86.28 & 76.11 & 18.71 & 4.63 \\
    SASSNet \cite{li2020shape} \textcolor{gray}{(MICCAI 2020)} & \multicolumn{1}{c}{8(10\%)} & \multicolumn{1}{c}{72}  & 85.22 & 75.09 & 11.18 & 2.89 \\
    DTC \cite{luo2021semi} \textcolor{gray}{(AAAI 2021)}   & \multicolumn{1}{c}{8(10\%)} & \multicolumn{1}{c}{72}  & {87.51} & {78.17} & {8.23} & 2.36 \\
    URPC \cite{luo2021efficient} \textcolor{gray}{(MICCAI 2021)}  & \multicolumn{1}{c}{8(10\%)} & \multicolumn{1}{c}{72} & 85.01 & 74.36 & 15.37 & 3.96 \\
    MRNet\cite{ji2021learning} \textcolor{gray}{(CVPR 2021)} &  \multicolumn{1}{c}{8(10\%)} & \multicolumn{1}{c}{72}  & 86.07 & 75.86 &  19.24 & 5.51 \\
    SCC \cite{liu2022contrastive} \textcolor{gray}{(CMIG 2022)}  & \multicolumn{1}{c}{8(10\%)} & \multicolumn{1}{c}{72} & 	86.51 & 76.54 & 10.51 & 2.56 \\
    ICT \cite{verma2022interpolation} \textcolor{gray}{(Neural Networks 2022)} &\multicolumn{1}{c}{8(10\%)} & \multicolumn{1}{c}{72} & 85.39 & 74.84  & 17.45 & 2.88 \\
    MC-Net+ \cite{wu2021semi} \textcolor{gray}{(MedIA 2022)} & \multicolumn{1}{c}{8(10\%)} & \multicolumn{1}{c}{72}  & 87.50  & 77.98 & 11.28 &{2.30} \\ 
    CauSSL\cite{miao2023caussl}  \textcolor{gray}{(ICCV 2023)} &  \multicolumn{1}{c}{8(10\%)} & \multicolumn{1}{c}{72}  & 87.49 & 77.95 & 18.85 & 5.11 \\
    PLGCL\cite{basak2023pseudo} \textcolor{gray}{(ICCV 2023)} &  \multicolumn{1}{c}{8(10\%)} & \multicolumn{1}{c}{72}  & 87.28 & 77.54 & 18.98 & 5.32 \\
    AUSS\cite{adiga2024anatomically} \textcolor{gray}{(MedIA 2024)} & \multicolumn{1}{c}{8(10\%)} & \multicolumn{1}{c}{72} & 87.57 & 78.32 & 8.17 & 2.22 \\
    MiDSS\cite{ma2024constructing} \textcolor{gray}{(CVPR 2024)} & \multicolumn{1}{c}{8(10\%)} & \multicolumn{1}{c}{72} & \textcolor{red}{88.03} & \textcolor{red}{78.69} & 8.04 & 2.16\\
    GALoss\cite{qi2024gradient} \textcolor{gray}{(ECCV 2024)} & \multicolumn{1}{c}{8(10\%)} & \multicolumn{1}{c}{72} & 87.86 & 78.55 & \textcolor{red}{8.01} & \textcolor{red}{2.07}\\
    \rowcolor{mygray}
    \textbf{SASNet} & \multicolumn{1}{c}{8(10\%)} & \multicolumn{1}{c}{72}  & \textbf{89.62} & \textbf{81.33} & \textbf{6.59} & \textbf{1.89} \\\hline
    UA-MT \cite{yu2019uncertainty} \textcolor{gray}{(MICCAI 2019)} &\multicolumn{1}{c}{16(20\%)} & \multicolumn{1}{c}{64}& 88.74 & 79.94 & 8.39  & 2.32 \\
    SASSNet \cite{li2020shape} \textcolor{gray}{(MICCAI 2020)} & \multicolumn{1}{c}{16(20\%)} & \multicolumn{1}{c}{64} & 89.16 & 80.60  & 8.95  & 2.26 \\
    DTC \cite{luo2021semi} \textcolor{gray}{(AAAI 2021)}   &  \multicolumn{1}{c}{16(20\%)} & \multicolumn{1}{c}{64} & 89.52 & 81.22 & 7.07  & 1.96 \\
    URPC \cite{luo2021efficient} \textcolor{gray}{(MICCAI 2021)} &  \multicolumn{1}{c}{16(20\%)} & \multicolumn{1}{c}{64}  & 88.74 & 79.93 & 12.73 & 3.66 \\
    MRNet\cite{ji2021learning} \textcolor{gray}{(CVPR 2021)} &  \multicolumn{1}{c}{16(20\%)} & \multicolumn{1}{c}{64}  & 88.62 & 80.94 & 8.83 & 2.48 \\
    SCC \cite{liu2022contrastive} \textcolor{gray}{(CMIG 2022)}  &  \multicolumn{1}{c}{16(20\%)} & \multicolumn{1}{c}{64} & 	89.81 & 81.64 & 7.15 & {1.82} \\
    ICT \cite{verma2022interpolation} \textcolor{gray}{(Neural Networks 2022)} &\multicolumn{1}{c}{16(20\%)} & \multicolumn{1}{c}{64}& 89.02 & 80.34  & 10.38 & 1.97 \\
    MC-Net+ \cite{wu2021semi} \textcolor{gray}{(MedIA 2022)} &   \multicolumn{1}{c}{16(20\%)} & \multicolumn{1}{c}{64}  & {90.12} & {82.12} & 8.07  & 1.99 \\
    CauSSL\cite{miao2023caussl} \textcolor{gray}{(ICCV 2023)} & \multicolumn{1}{c}{16(20\%)} & \multicolumn{1}{c}{64}   & 90.16 & 82.17 & \textcolor{red}{6.11}   & 1.97 \\
    PLGCL\cite{basak2023pseudo} \textcolor{gray}{(ICCV 2023)} &  \multicolumn{1}{c}{16(20\%)} & \multicolumn{1}{c}{64}  & 90.01 &  82.04& 6.21 & 2.15 \\
    AUSS\cite{adiga2024anatomically} \textcolor{gray}{(MedIA 2024)} &  \multicolumn{1}{c}{16(20\%)} & \multicolumn{1}{c}{64} & \textcolor{red}{90.79} & \textcolor{red}{82.91} & 6.13 & 1.93 \\
    MiDSS\cite{ma2024constructing} \textcolor{gray}{(CVPR 2024)} & \multicolumn{1}{c}{16(20\%)} & \multicolumn{1}{c}{64} & 90.47 & 82.55 & 6.17 & \textcolor{red}{1.80}\\
    GALoss\cite{qi2024gradient} \textcolor{gray}{(ECCV 2024)} & \multicolumn{1}{c}{16(20\%)} & \multicolumn{1}{c}{64} & 90.39 & 82.46 & 6.35 & 2.01\\
    \rowcolor{mygray}
    \textbf{SASNet} &  \multicolumn{1}{c}{16(20\%)} & \multicolumn{1}{c}{64} & \textbf{91.82} & \textbf{84.93} & \textbf{4.63} & \textbf{1.42} \\
    \bottomrule
    \end{tabular}}
    \caption{Compared with the other SOTA methods on the LA dataset with 10\% and 20\% of the labeled data. \textbf{Bold} represents the best performance, while \textcolor{red}{red} represents the second-best performance.}
    \label{tab:LA_result_table}%
    \vspace{-4mm}
\end{table*}%

\subsection{Implementation Details}
All our experiments are conducted in Pytorch 1.12 and CUDA 11.3 framework with NVIDIA A100 GPU. Our proposed SASNet employs V-Net as the backbone. In the training phase, we randomly crop to a size of $112\times112\times80$ for the LA dataset, and we randomly crop to a size of $96\times96\times96$ for the Pancreas-CT dataset and the BraTS dataset. The SGD optimizer is utilized with an initial learning rate of 0.01. The batch size is 4, and the temperature in the sharpening function is 0.1. In the testing phase, we use Dice coefficient (Dice), Jaccard coefficient (Jaccard), Average surface distance (ASD) and 95\% Hausdorff Distance (HD95) to compare the performance difference with other methods.

\begin{table*}[!ht] 
\footnotesize
  \centering
  \resizebox{0.85\columnwidth}{!}{%
    \begin{tabular}{crrcccc}
    \toprule
    \multirow{2}[0]{*}{Method} & \multicolumn{2}{c}{\#Scans used} & \multicolumn{4}{c}{Metrics} \\
    \cmidrule(r){2-3} \cmidrule(l){4-7} 
          & \multicolumn{1}{c}{Labeled} & \multicolumn{1}{c}{Un} & Dice(\%) $\uparrow$ & Jaccard(\%) $\uparrow$ & HD95 (voxel) $\downarrow$ & ASD (voxel) $\downarrow$ \\
    \midrule
    V-Net\cite{milletari2016v} \textcolor{gray}{(3DV 2016)} & \multicolumn{1}{c}{6(10\%)} & \multicolumn{1}{c}{0} & 54.94 & 40.87 & 47.48 & 17.43 \\
    V-Net\cite{milletari2016v} \textcolor{gray}{(3DV 2016)} & \multicolumn{1}{c}{12(20\%)} & \multicolumn{1}{c}{0} & 71.52 & 57.68 & 18.12 & 5.41 \\
    V-Net\cite{milletari2016v} \textcolor{gray}{(3DV 2016)} & \multicolumn{1}{c}{62(100\%)} & \multicolumn{1}{c}{0} & 82.60  & 70.81 & 5.61  & 1.33 \\\hline
    UA-MT\cite{yu2019uncertainty} \textcolor{gray}{(MICCAI 2019)} & \multicolumn{1}{c}{6(10\%)} & \multicolumn{1}{c}{56} & 66.44 & 52.02 & 17.04 & 3.03 \\
    SASSNet\cite{li2020shape} \textcolor{gray}{(MICCAI 2020)} & \multicolumn{1}{c}{6(10\%)} & \multicolumn{1}{c}{56}  & 68.97 & 54.29 & 18.83 & \textcolor{red}{1.96} \\
    DTC\cite{luo2021semi}  \textcolor{gray}{(AAAI 2021)}   & \multicolumn{1}{c}{6(10\%)} & \multicolumn{1}{c}{56}  & 66.27 & 52.07 & 15.00 & 4.44 \\
    URPC\cite{luo2021efficient} \textcolor{gray}{(MICCAI 2021)}  &  \multicolumn{1}{c}{6(10\%)} & \multicolumn{1}{c}{56} & 73.53 & 59.44 & 22.57 & 7.85 \\
    MRNet\cite{ji2021learning}  \textcolor{gray}{(CVPR 2021)} &  \multicolumn{1}{c}{6(10\%)} & \multicolumn{1}{c}{56}  & 72.47 & 56.92 & 15.08 & 5.14 \\
    MC-Net+\cite{wu2021semi} \textcolor{gray}{(MedIA 2022)} &  \multicolumn{1}{c}{6(10\%)} & \multicolumn{1}{c}{56}  & 68.94 & 54.74 & 16.28 & 3.16 \\  
    PLGCL\cite{basak2023pseudo} \textcolor{gray}{(CVPR 2023)} &  \multicolumn{1}{c}{6(10\%)} & \multicolumn{1}{c}{56}  & 73.15 & 58.83 & 14.36 & 4.51 \\
    BCP\cite{bai2023bidirectional}  \textcolor{gray}{(CVPR 2023)} &  \multicolumn{1}{c}{6(10\%)} & \multicolumn{1}{c}{56} & \textcolor{red}{74.25} & \textcolor{red}{60.03} & 14.23 & 4.42 \\
    AUSS\cite{adiga2024anatomically} \textcolor{gray}{(MedIA 2024)} &  \multicolumn{1}{c}{6(10\%)} & \multicolumn{1}{c}{56} & 73.82 & 59.33 & 14.68 & 4.95\\
    MiDSS\cite{ma2024constructing} \textcolor{gray}{(CVPR 2024)} &  \multicolumn{1}{c}{6(10\%)} & \multicolumn{1}{c}{56} & 74.12 & 59.84 & \textcolor{red}{13.93} & {3.97}\\
    GALoss\cite{qi2024gradient} \textcolor{gray}{(ECCV 2024)} &  \multicolumn{1}{c}{6(10\%)} & \multicolumn{1}{c}{56} & 73.94 & 59.51 & 14.52 & 4.71\\
    \rowcolor{mygray}
    \textbf{SASNet} &  \multicolumn{1}{c}{6(10\%)} & \multicolumn{1}{c}{56} & \textbf{76.38} & \textbf{62.84} & \textbf{13.47} & \textbf{1.82} \\\hline
    UA-MT\cite{yu2019uncertainty} \textcolor{gray}{(MICCAI 2019)} & \multicolumn{1}{c}{12(20\%)} & \multicolumn{1}{c}{50} & 76.10  & 62.62 & 10.84 & 2.43 \\
    SASSNet\cite{li2020shape} \textcolor{gray}{(MICCAI 2020)} &  \multicolumn{1}{c}{12(20\%)} & \multicolumn{1}{c}{50}   & 76.39 & 63.17 & 11.06 & \textbf{1.42} \\
    DTC\cite{luo2021semi} \textcolor{gray}{(AAAI 2021)}    &  \multicolumn{1}{c}{12(20\%)} & \multicolumn{1}{c}{50}   & 78.27 & 64.75 & \textbf{8.36} & 2.25 \\   
    URPC\cite{luo2021efficient} \textcolor{gray}{(MICCAI 2021)} &  \multicolumn{1}{c}{12(20\%)} & \multicolumn{1}{c}{50}   & 80.02 & 67.30  & \textcolor{red}{8.51}  & 1.98 \\
    MRNet\cite{ji2021learning}  \textcolor{gray}{(CVPR 2021)} &  \multicolumn{1}{c}{12(20\%)} & \multicolumn{1}{c}{50}  & 77.82 & 63.85 & 13.86 & 4.53 \\
    MC-Net+\cite{wu2021semi} \textcolor{gray}{(MedIA 2022)} & \multicolumn{1}{c}{12(20\%)} & \multicolumn{1}{c}{50}   & 79.05 & 65.83 & {10.29}   & {2.72} \\  
    PLGCL\cite{basak2023pseudo} \textcolor{gray}{(CVPR 2023)} &  \multicolumn{1}{c}{12(20\%)} & \multicolumn{1}{c}{50}  & 78.41 & 65.17 & 14.13 & 4.68 \\
    BCP\cite{bai2023bidirectional} \textcolor{gray}{(CVPR 2023)} &  \multicolumn{1}{c}{12(20\%)} & \multicolumn{1}{c}{50}   & \textcolor{red}{80.37} & \textcolor{red}{67.81} & 11.53 & {2.06} \\
    AUSS\cite{adiga2024anatomically} \textcolor{gray}{(MedIA 2024)} &  \multicolumn{1}{c}{12(20\%)} & \multicolumn{1}{c}{50} & 80.02 & 66.73 & 12.38 & 3.14 \\
    MiDSS\cite{ma2024constructing} \textcolor{gray}{(CVPR 2024)} &  \multicolumn{1}{c}{12(20\%)} & \multicolumn{1}{c}{50} & 79.74 & 66.56 & 12.65 & 3.47 \\
    GALoss\cite{qi2024gradient} \textcolor{gray}{(ECCV 2024)} &  \multicolumn{1}{c}{12(20\%)} & \multicolumn{1}{c}{50} & 80.21 & 66.92 & 11.78 & 2.83 \\
    \rowcolor{mygray}
    \textbf{SASNet} &  \multicolumn{1}{c}{12(20\%)} & \multicolumn{1}{c}{50}   & \textbf{81.60} & \textbf{69.39} & 11.25 & \textcolor{red}{1.81} \\
    \bottomrule
    \end{tabular}}
    \caption{Compared with the other SOTA methods on the Pancreas-CT dataset with 10\% and 20\% of the labeled data. \textbf{Bold} represents the best performance, while \textcolor{red}{red} represents the second-best performance.}
    \label{tab:Pancreas_result_table}
    \vspace{-4mm}
\end{table*}

\subsection{Performance Comparison with Other Methods}

\subsubsection{Performance on LA Dataset}
Table \ref{tab:LA_result_table} presents a comparison of our method with other state-of-the-art approaches on the LA dataset. We evaluate our method on the V-Net using 10\% and 20\% labeled data as well as fully supervised data.
With only 10\% labeled data at hand, our model exhibits noteworthy enhancements over DTC \cite{luo2021semi}. Specifically, we achieve improvements of 2.11\% and 3.16\% in Dice and Jaccard scores, respectively. When 20\% labeled data is available, our method achieves a 1.70\% improvement in Dice and a 2.81\% improvement in Jaccard compared to MC-Net+\cite{wu2021semi}. Notably, our approach even outperforms the fully supervised results obtained on V-Net.
Based on the visualizations in Fig. \ref{LA_Pancreas_result}, the first and second rows demonstrate that with only 20\% labeled data, other baseline models exhibit fragmentation or missing portions in the challenging areas. In contrast, our method achieves comprehensive segmentation in these challenging regions.

\subsubsection{Performance on Pancreas-CT Dataset}
Table \ref{tab:Pancreas_result_table} presents a comparison of our method with other state-of-the-art techniques on the Pancreas-CT dataset. When the labeled data is limited to only 10\%, SASNet achieves significant improvements compared to other methods. 
Specifically, our method outperforms URPC \cite{luo2021efficient}  in Dice score by 2.85\% and in Jaccard by 3.40\%. As the percentage of labeled data increased to 20\%, our method outperforms URPC \cite{luo2021efficient}  in Dice score by 1.58\% and in Jaccard score by 2.09\%.
The third and fourth rows of Fig. \ref{LA_Pancreas_result} illustrate the visual results on the Pancreas-CT dataset with 20\% labeled data, other methods tend to misclassify background as foreground. In contrast, our method effectively addresses this issue and accurately segments the target.

\begin{table*}[!t] 
\footnotesize
  \centering
  \resizebox{0.85\columnwidth}{!}{%
    \begin{tabular}{crrcccc}
    \toprule
    \multirow{2}[0]{*}{Method} & \multicolumn{2}{c}{\#Scans used} & \multicolumn{4}{c}{Metrics} \\
    \cmidrule(r){2-3} \cmidrule(l){4-7} 
          & \multicolumn{1}{c}{Labeled} & \multicolumn{1}{c}{Un} & Dice(\%) $\uparrow$ & Jaccard(\%) $\uparrow$ & HD95 (voxel) $\downarrow$ & ASD (voxel) $\downarrow$ \\
    \midrule
    V-Net\cite{milletari2016v} \textcolor{gray}{(3DV 2016)} & \multicolumn{1}{c}{25(10\%)} & \multicolumn{1}{c}{0} & 73.00 & 59.96 & 42.64 & 2.96 \\
    V-Net\cite{milletari2016v} \textcolor{gray}{(3DV 2016)} & \multicolumn{1}{c}{50(20\%)} & \multicolumn{1}{c}{0} & 76.14 & 64.15 & 36.01 & 2.70 \\
    V-Net\cite{milletari2016v} \textcolor{gray}{(3DV 2016)} & \multicolumn{1}{c}{250(100\%)} & \multicolumn{1}{c}{0} & 84.81  & 75.53 & 8.08  & 1.94 \\\hline
    EM\cite{vu2019advent} \textcolor{gray}{(ECCV 2019)} & \multicolumn{1}{c}{25(10\%)} & \multicolumn{1}{c}{225} & 81.74 & 71.42  &13.71 & {2.37} \\
    UA-MT\cite{yu2019uncertainty} \textcolor{gray}{(MICCAI 2019)} & \multicolumn{1}{c}{25(10\%)} & \multicolumn{1}{c}{225} & 80.85 & 70.32 & 14.61 & 2.57 \\
    DTC\cite{luo2021semi} \textcolor{gray}{(AAAI 2021)}    & \multicolumn{1}{c}{25(10\%)} & \multicolumn{1}{c}{225}  & {81.96} & {71.84} & 12.08 & 2.43 \\ 
    URPC\cite{luo2021efficient} \textcolor{gray}{(MICCAI 2021)}  &  \multicolumn{1}{c}{25(10\%)} & \multicolumn{1}{c}{225} & 81.80 & 71.63 & {11.50} & 2.48 \\
    AUSS\cite{adiga2024anatomically} \textcolor{gray}{(MedIA 2024)} & \multicolumn{1}{c}{25(10\%)} & \multicolumn{1}{c}{225} & 81.99 & 71.92 & 11.47 & 2.44\\
    MiDSS\cite{ma2024constructing} \textcolor{gray}{(CVPR 2024)} & \multicolumn{1}{c}{25(10\%)} & \multicolumn{1}{c}{225} & 81.91 & 71.03 & 11.38 &  \textcolor{red}{2.37}\\
    GALoss\cite{qi2024gradient} \textcolor{gray}{(ECCV 2024)} & \multicolumn{1}{c}{25(10\%)} & \multicolumn{1}{c}{225} & \textcolor{red}{82.13} & \textcolor{red}{72.11} & \textcolor{red}{11.25} & 2.39\\
    \rowcolor{mygray}
    \textbf{SASNet} &  \multicolumn{1}{c}{25(10\%)} & \multicolumn{1}{c}{225} & \textbf{82.84} & \textbf{73.00} & \textbf{10.91} & \textbf{2.31} \\\hline
    EM\cite{vu2019advent} \textcolor{gray}{(ECCV 2019)} &\multicolumn{1}{c}{50(20\%)} & \multicolumn{1}{c}{200} & 82.37 & 72.28  & 15.83 & 2.30 \\
    UA-MT\cite{yu2019uncertainty} \textcolor{gray}{(MICCAI 2019)} & \multicolumn{1}{c}{50(20\%)} & \multicolumn{1}{c}{200} & 81.87  & 71.42 & 13.98 & 2.49 \\
    DTC\cite{luo2021semi} \textcolor{gray}{(AAAI 2021)}    &  \multicolumn{1}{c}{50(20\%)} & \multicolumn{1}{c}{200}   & 82.78 & 72.47 & 13.43 & {2.20} \\    
    URPC\cite{luo2021efficient} \textcolor{gray}{(MICCAI 2021)} &  \multicolumn{1}{c}{50(20\%)} & \multicolumn{1}{c}{200}   & {82.80} & {72.72}  & {12.48}  & 2.72 \\
    AUSS\cite{adiga2024anatomically} \textcolor{gray}{(MedIA 2024)} &  \multicolumn{1}{c}{50(20\%)} & \multicolumn{1}{c}{200} & 83.21 & 73.55 & 11.39 & \textcolor{red}{2.03} \\
    MiDSS\cite{ma2024constructing} \textcolor{gray}{(CVPR 2024)} &  \multicolumn{1}{c}{50(20\%)} & \multicolumn{1}{c}{200} & \textcolor{red}{83.74} & \textcolor{red}{73.93} & 11.06 & 2.29\\
    GALoss\cite{qi2024gradient} \textcolor{gray}{(ECCV 2024)} &  \multicolumn{1}{c}{50(20\%)} & \multicolumn{1}{c}{200} & 83.49 & 73.72 & \textcolor{red}{10.61} & 2.18\\
    \rowcolor{mygray}
    \textbf{SASNet} &  \multicolumn{1}{c}{50(20\%)} & \multicolumn{1}{c}{200}   & \textbf{85.84} & \textbf{76.79} & \textbf{7.52} & \textbf{1.62} \\
    \bottomrule
    \end{tabular}}
    \caption{Compared with the other SOTA methods on the BraTS dataset with 10\% and 20\% of the labeled data. \textbf{Bold} represents the best performance, while \textcolor{red}{red} represents the second-best performance.}
    \label{tab:BraTS19_result_table}%
    \vspace{-4mm}
\end{table*}

\subsubsection{Performance on BraTS Dataset}
Table \ref{tab:BraTS19_result_table} compares our method with other state-of-the-art approaches on the BraTS dataset. Notably, our method achieves superior performance even with only 20\% of the data labeled, outperforming fully supervised approaches. Additionally, when compared to other semi-supervised methods with the same amount of labeling, such as URPC \cite{luo2021efficient}, our method surpasses it by 3.04\% in Dice score and by 4.07\% in Jaccard score. \textcolor{Revision}{ As shown in Fig. \ref{BraTs_result}, in the first example, other methods misclassify the background as foreground in areas where the lesion boundaries are difficult to distinguish. In the second example, while other methods erroneously predict a non-lesion region within the lesion area as part of the lesion, our model successfully avoids this misclassification, demonstrating superior segmentation capability in terms of finer details.}

\subsection{Ablation Experiments}
\subsubsection{Effects of different branch numbers.}
In the absence of SAR, we evaluate the proposed model with different numbers of branches, ranging from one to two, and compare their performance using three different loss functions: $L_{seg}$, $L_{src}$, and $L_{plc}$. Specifically, In Table \ref{tab:LA_branch}, we observe that the two-branch model outperforms the one-branch model (high-level) in terms of the Dice coefficient when only $L_{seg}$ is used or when $L_{seg}$ and $L_{src}$ are used simultaneously. For instance, on the LA dataset, the Dice coefficient of the two-branch model is 0.51\% higher than that of the one-branch MC-Net+ \cite{wu2021semi} when only $L_{seg}$ is used. 
The result suggests that the additional low-level branch helps to improve the segmentation performance by leveraging the complementary information in different loss functions. 

\begin{table*}[ht]
 \footnotesize
  \centering
  \resizebox{0.9\linewidth}{!}{%
    \begin{tabular}{c|cccc|cccc}
    \toprule
        Method  &   Nums    & $L_{seg}$ & $L_{src}$ & $L_{plc}$ &\multicolumn{1}{l}{Dice (\%) $\uparrow$} & \multicolumn{1}{c}{Jaccard (\%) $\uparrow$} & \multicolumn{1}{c}{HD95 (voxel) $\downarrow$} & \multicolumn{1}{c}{ASD (voxel) $\downarrow$} \\
    \midrule
    {MC-Net+} & 1     & \checkmark    &  & & 90.12 & 82.12 & 8.07  & 1.99 \\
    SASNet& 1     & \checkmark    &  \multicolumn{1}{c}{\checkmark}  &  & 90.23 & 82.37 & 8.90   & 2.44 \\ \midrule%
      SASNet    & 2     & \checkmark      & & \multicolumn{1}{c}{} & 90.63 & 82.99 & 6.50  & 2.13 \\%
      SASNet& 2     & \checkmark      & \multicolumn{1}{c}{\checkmark} &      & 90.94 & 83.85  & 6.01 & 1.93\\%
      SASNet& 2     & \checkmark      & \checkmark      & \checkmark      & \textbf{91.52} & \textbf{84.42} & \textbf{4.98} & \textbf{1.48} \\%
          \bottomrule
    \end{tabular}}
    \caption{Results of different branch numbers and training objectives on the LA dataset.}
    \label{tab:LA_branch}%
\end{table*}

\begin{table*}[!htbp]\footnotesize
  \centering
  \resizebox{\linewidth}{!}{%
    \begin{tabular}{c|cccc|cccc}
    \toprule
    Method & View Variance  & $L_{seg}$ & $L_{src}$ & $L_{plc}$  &\multicolumn{1}{l}{Dice(\%) $\uparrow$} & \multicolumn{1}{l}{Jaccard(\%) $\uparrow$} & \multicolumn{1}{l}{HD95 (voxel)$\downarrow$} & \multicolumn{1}{l}{ASD (voxel)$\downarrow$} \\
          \midrule
    SASNet    & \multicolumn{1}{c}{\checkmark} & \checkmark     &       &       & 90.63 & 82.99 & 6.50  & 2.13 \\%
    SASNet    & \multicolumn{1}{c}{\checkmark} & \checkmark     & \checkmark     &      & 90.94 &  83.53     & 6.01      & 1.93\\ %
    SASNet    & \multicolumn{1}{c}{\checkmark} & \checkmark     &       & \checkmark  & 91.25 & 84.00 & 5.06  & 1.56 \\%
    SASNet    & & \checkmark     & \checkmark     & \checkmark   & 91.16 & 83.84 & 5.80   & 1.65 \\
    SASNet    & \multicolumn{1}{c}{\checkmark} & \checkmark     & \checkmark     & \checkmark    & \textbf{91.52} & \textbf{84.42} &\textbf{4.98}  & \textbf{1.48}\\
    \bottomrule
    \end{tabular}%
    }  
\caption{Results of view variance enhancement mechanism and training objectives without SAR on the LA dataset.}
\label{tab:LA_em}%
\end{table*}%

\subsubsection{Effects of enhancement mechanism and training objectives.}
Table \ref{tab:LA_em} displays ablation results for both the augmentation mechanism and training objectives without SAR on the LA dataset. It is evident from the table that the inclusion of either $L_{src}$ or $L_{plc}$ yields performance improvements in the presence of the augmentation mechanism, surpassing the performance achieved by using only $L_{seg}$. This phenomenon serves as compelling evidence of the efficacy of our designed loss functions.
Furthermore, the model's performance gains further momentum when all three loss functions are employed. Additionally, a comparison between the results showcased in the fourth and fifth rows vividly illustrates the effectiveness of our proposed view variance mechanism. 

\subsubsection{Ablation studies of SAR/SDM/low-level on the datasets.} 
\textcolor{Revision}{We conducted comprehensive ablation experiments across all three evaluated datasets to systematically assess the individual contributions of our proposed Scale-aware Adaptive Reweight mechanism, Signed Distance Map processing, and low-level feature integration. As shown in Table \ref{tab:SSl_1}-\ref{tab:SSl_3}, the experimental results indicate that incorporating SAR, SDM, and low-level methods into the model can significantly improve performance. The Scale-aware Adaptive Reweight strategy yields substantial improvements across all datasets, with Dice coefficient enhancements of 0.30\% on LA, 0.74\% on Pancreas-CT, and 0.27\% on BraTS datasets, accompanied by consistent boundary precision improvements reflected in reduced HD95 distances. These findings substantiate the effectiveness of our confidence-based pixel-wise weighting approach in generating more reliable ensemble predictions by selectively emphasizing high-confidence regions while mitigating potential errors from unreliable predictions. The Signed Distance Map component demonstrates the robust performance gains, achieving Dice coefficient improvements of 0.92\% on LA, 0.93\% on Pancreas-CT, and 0.52\% on BraTS datasets, which facilitates effective segmentation-regression consistency learning that bridges geometric understanding with probabilistic predictions. The integration of low-level features consistently enhances segmentation performance across all evaluated datasets, with notable improvements in small structure detection and boundary refinement, confirming that the complementary utilization of multi-scale feature representations enables more comprehensive anatomical understanding. The consistent performance patterns across heterogeneous datasets spanning cardiac, abdominal, and neurological imaging modalities demonstrate the broad applicability and methodological robustness of our proposed components, establishing their importance in advancing semi-supervised medical image segmentation under limited annotation scenarios.}

\begin{table*}[!ht]
  \centering
  \resizebox{0.7\columnwidth}{!}{%
    \begin{tabular}{c|cccc}
    \toprule
       \multirow{2}*{Method} & \multicolumn{4}{c}{LA Dataset} \\
       \cmidrule(lr){2-5}
        & \multicolumn{1}{c}{Dice(\%) $\uparrow$} & \multicolumn{1}{c}{Jaccard(\%) $\uparrow$} & \multicolumn{1}{c}{HD95 (voxel) $\downarrow$} & \multicolumn{1}{c}{ASD (voxel) $\downarrow$} \\
          \midrule
    w/o SAR & 91.52 & 84.42 & 4.98  & 1.48 \\
    with SAR  & 91.82 \textcolor{red}{(0.30\% $\uparrow$)} & 84.93 \textcolor{red}{(0.51\% $\uparrow$)} & 4.63 \textcolor{red}{(0.35 $\downarrow$)} & 1.42 \textcolor{red}{(0.06 $\downarrow$)}\\
    \midrule
    w/o SDM & 90.90 & 83.39 & 5.13 & 1.57 \\
    with SDM  & 91.82 \textcolor{red}{(0.92\% $\uparrow$)} & 84.93 \textcolor{red}{(1.54\% $\uparrow$)}& 4.63 \textcolor{red}{(0.50 $\downarrow$)}& 1.42 \textcolor{red}{(0.15 $\downarrow$)}\\
    \midrule
   w/o low-level   & 90.23 & 82.37 & 8.90   & 2.44 \\
    with low-level   & 90.94 \textcolor{red}{(0.71\% $\uparrow$)}& 83.85 \textcolor{red}{(1.48\% $\uparrow$)}& 6.01 \textcolor{red}{(2.89 $\downarrow$)}& 1.93 \textcolor{red}{(0.51 $\downarrow$)}\\
    \bottomrule
    \end{tabular}}
      \caption{Ablation studies of SAR/SDM/low-level on the LA dataset.}     
  \label{tab:SSl_1}%
  \vspace{-4mm}
\end{table*}%

\begin{table*}[!ht]
  \centering
  \resizebox{0.7\columnwidth}{!}{%
    \begin{tabular}{c|cccc}
    \toprule
       \multirow{2}*{Method} & \multicolumn{4}{c}{Pancreas-CT Dataset} \\
       \cmidrule(lr){2-5} 
        & \multicolumn{1}{c}{Dice(\%) $\uparrow$} & \multicolumn{1}{c}{Jaccard(\%) $\uparrow$} & \multicolumn{1}{c}{HD95 (voxel) $\downarrow$} & \multicolumn{1}{c}{ASD (voxel) $\downarrow$} \\
          \midrule
    w/o SAR & 80.86 & 68.49 & 8.55 & 1.31 \\
    with SAR  & 81.60 \textcolor{red}{(0.74\% $\uparrow$)}& 69.39 \textcolor{red}{(0.90\% $\uparrow$)}& 11.25 \textcolor{red}{(2.70 $\uparrow$)}& 1.81 \textcolor{red}{(0.50 $\uparrow$)}\\
    \midrule
    w/o SDM & 80.67 & 68.13 & 12.72 & 1.77 \\
    with SDM  & 81.60 \textcolor{red}{(0.93\% $\uparrow$)}& 69.39 \textcolor{red}{(1.26\% $\uparrow$)}& 11.25 \textcolor{red}{(1.47 $\downarrow$)}& 1.81 \textcolor{red}{(0.04 $\uparrow$)}\\
    \midrule
   w/o low-level  & 79.56 & 66.84 & 10.41   & 1.62\\
    with low-level & 80.22 \textcolor{red}{(0.66\% $\uparrow$)}& 67.59 \textcolor{red}{(0.75\% $\uparrow$)}&10.20 \textcolor{red}{(0.21 $\downarrow$)}& 1.92 \textcolor{red}{(0.30 $\uparrow$)}\\
    \bottomrule
    \end{tabular}%
  }
      \caption{Ablation studies of SAR/SDM/low-level on the Pancreas-CT dataset.}     
  \label{tab:SSl_2}%
\end{table*}%

\begin{table*}[!ht]
  \centering
  \resizebox{0.7\columnwidth}{!}{%
    \begin{tabular}{c|cccc}
    \toprule
       \multirow{2}*{Method} & \multicolumn{4}{c}{BraTS Dataset} \\
       \cmidrule(lr){2-5} 
        & \multicolumn{1}{c}{Dice(\%) $\uparrow$} & \multicolumn{1}{c}{Jaccard(\%) $\uparrow$} & \multicolumn{1}{c}{HD95 (voxel) $\downarrow$} & \multicolumn{1}{c}{ASD (voxel) $\downarrow$} \\
          \midrule
    w/o SAR & 85.57 & 76.33 & 7.73 & 1.81 \\
    with SAR  & 85.84 \textcolor{red}{(0.27\% $\uparrow$)}& 76.79 \textcolor{red}{(0.46\% $\uparrow$)}& 7.52 \textcolor{red}{(0.21 $\downarrow$)}& 1.62 \textcolor{red}{(0.19 $\downarrow$)}\\
    \midrule 
    w/o SDM & 85.32 & 76.16 & 7.89 & 1.90 \\
    with SDM  & 85.84 \textcolor{red}{(0.52\% $\uparrow$)}& 76.79 \textcolor{red}{(0.63\% $\uparrow$)}& 7.52 \textcolor{red}{(0.37 $\downarrow$)}& 1.62 \textcolor{red}{(0.28 $\downarrow$)}\\
    \midrule
    w/o low-level & 84.39 & 75.22 & 8.67 & 2.12 \\
    with low-level & 85.03 \textcolor{red}{(0.64\% $\uparrow$)}& 75.94 \textcolor{red}{(0.72\% $\uparrow$)}& 8.25 \textcolor{red}{(0.42 $\downarrow$)}& 1.96 \textcolor{red}{(0.16 $\downarrow$)}\\
    \bottomrule
    \end{tabular}%
  }
      \caption{\textcolor{Revision}{Ablation studies of SAR/SDM/low-level on the BraTS dataset.}}     
  \label{tab:SSl_3}%
\end{table*}%

\subsubsection{Effects of different layers as the low-level decoder.} 
Table \ref{tab:diff_low-level} presents the ablation results for different numbers of layers in the low-level decoder. It is clear that the 4-layer decoder consistently outperforms the 3-layer configuration across all evaluation metrics. This indicates that a 4-layer low-level decoder more effectively captures fine-grained local features, which is particularly important for tasks requiring precise boundary delineation. The concurrent improvement in both region-based and boundary-based metrics suggests that increasing the depth of the low-level decoder not only preserves high-level semantic information but also enhances the representation of local spatial details. By comparison, the 3-layer decoder, while still effective, does not fully leverage the rich local features provided by the encoder. These observations underscore the strong compatibility and efficacy of the 4-layer configuration as an adept low-level decoder.

\begin{table*}[!htbp]\scriptsize
  \label{tab:diff_low-level}%
  \centering
  \resizebox{0.7\columnwidth}{!}{%
    \begin{tabular}{c|cccc}
    \toprule
       Method  & \multicolumn{1}{c}{Dice(\%) $\uparrow$} & \multicolumn{1}{c}{Jaccard(\%) $\uparrow$} & \multicolumn{1}{c}{HD95 (voxel) $\downarrow$} & \multicolumn{1}{c}{ASD (voxel) $\downarrow$} \\
          \midrule
    SASNet (3 Layers) & 91.20 & 83.90 & 5.38 & 1.80 \\ %
    SASNet (4 Layers)  & \textbf{91.82} \textcolor{red}{(0.62\% $\uparrow$)}& \textbf{84.93} \textcolor{red}{(1.03\% $\uparrow$)}& \textbf{4.63} \textcolor{red}{(0.75 $\downarrow$)}& \textbf{1.42} \textcolor{red}{(0.38 $\downarrow$)}\\%
    \bottomrule
    \end{tabular}}
    \caption{Results of employing different layers to the low-level decoder on LA dataset.}
    \label{tab:diff_low-level}
    \vspace{-4mm}
\end{table*}%

\textcolor{Revision}{\subsubsection{Ablation Studies of Different Hyperparameters}}
\textcolor{Revision}{To establish robust methodological validation and enhance experimental reproducibility, we conducted ablation experiments investigating the impact of critical hyperparameters $\beta$ and $\gamma$ on segmentation performance across five distinct configurations using the LA dataset with 20\% labeled data as shown in Table \ref{tab:ablation_parameters}. The comprehensive analysis demonstrates several insights that validate our theoretical design choices. Direct application of $\gamma$ equal to one without ramp-up scheduling yields suboptimal performance with a Dice coefficient of 91.36\%, demonstrating the necessity of progressive consistency weight escalation during training initialization to prevent pseudo-label propagation instabilities that can destabilize early learning dynamics. The optimal configuration combining $\beta$ equal to 0.5 with $\gamma$ equal to one under sigmoid ramp-up scheduling achieves superior performance across all evaluation metrics, attaining a Dice coefficient of 91.82\%, Jaccard coefficient of 84.93\%, HD95 distance of 4.63 voxels, and average surface distance of 1.42 voxels. This configuration establishes the effectiveness of balanced supervised-consistency learning that equalizes the influence of labeled supervision and semi-supervised regularization components. Sensitivity analysis across different $\gamma$ values ranging from 0.5 to one under optimal $\beta$ conditions demonstrates that stronger consistency regularization enhances segmentation accuracy, while modifications to $\beta$ beyond the balanced configuration exhibit diminishing returns in performance gains.}

\begin{table*}[!htbp]\scriptsize
  \centering
  \resizebox{0.8\columnwidth}{!}{%
    \begin{tabular}{cc|cccc}
    \toprule
       \multicolumn{2}{c}{Hyperparameter}  & \multicolumn{1}{c}{Dice(\%) $\uparrow$} & \multicolumn{1}{c}{Jaccard(\%) $\uparrow$} & \multicolumn{1}{c}{HD95 (voxel) $\downarrow$} & \multicolumn{1}{c}{ASD (voxel) $\downarrow$} \\ \midrule
    $\beta$: 0.5 & $\gamma$: 1.0 (w/o ramp-up) & 91.36 & 84.24 & 5.02 & 1.75 \\ \midrule
    $\beta$: 0.5 & $\gamma$: 0.5 (with ramp-up) & 91.02 & 83.87 & 5.12 & 1.98 \\
    $\beta$: 0.5 & $\gamma$: 1.0 (with ramp-up) & \textbf{91.82} & \textbf{84.93} & \textbf{4.63} & \textbf{1.42} \\
    $\beta$: 1.0 & $\gamma$: 0.5 (with ramp-up) & 91.15 & 84.03 & 5.06 & 1.86\\
    $\beta$: 1.0 & $\gamma$: 1.0 (with ramp-up) & 91.64 & 84.55 & 4.91 & 1.47 \\
    \bottomrule
    \end{tabular}}
    \caption{\textcolor{Revision}{Ablation studies of the hyperparameters ($\beta$ and $\gamma$) on the LA datasets.}}
    \label{tab:ablation_parameters}
\end{table*}%

\subsection{Interpretability Analysis}
\begin{figure}[!htbp]
\centering
  \includegraphics[width=0.8\linewidth]{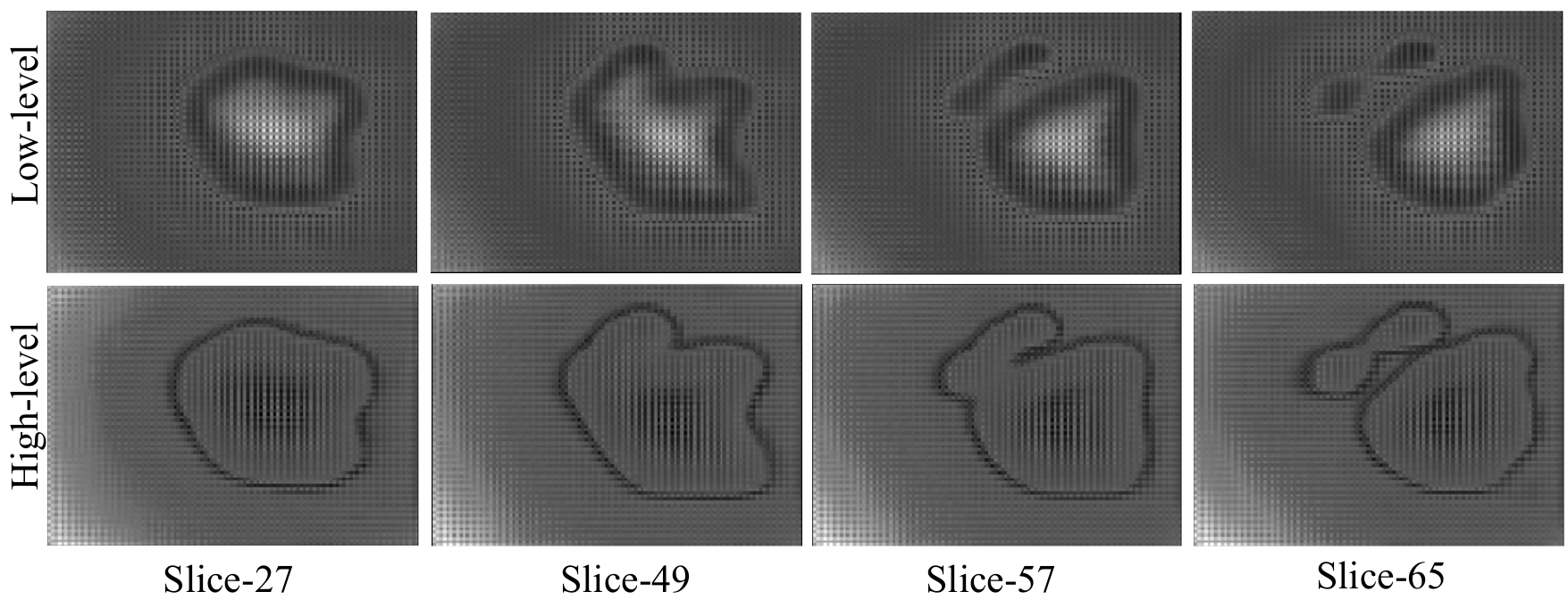}
  \caption{Visual comparison between low-level and high-level feature maps of same case.}
  \label{feature_map}
  \vspace{-2mm}
\end{figure}

Fig. \ref{feature_map} presents feature maps from the middle layer's convolutional blocks in the decoder. 
By contrasting the feature maps of high-level and low-level branches, we notice that the low-level branch pays greater attention to local features, aiding in distinguishing between foreground and background. In contrast, the high-level branch tends to capture more global and abstract features, that can provide a holistic understanding of the image. Additionally, we note that the high-level branches are equipped with more residual connections than the low-level branches, which allows the high-level branch to integrate fine-grained details from the lower layers and refine the boundaries between objects and background, resulting in more accurate and coherent segmentation maps. 
Overall, the multi-scale branch achieves a good balance between local and global information, and produces feature maps with high-quality boundary cues.

%\vspace{-4mm}

\subsection{\textcolor{Revision}{Model Complexity Analysis}}

\textcolor{Revision}{
As shown in Table~\ref{tab:complexity}, SASNet has a relatively large number of parameters, while its FLOPs are slightly lower than URPC\cite{luo2021efficient} . The increased complexity is mainly due to SASNet integrating the predictions from two branches to improve the final prediction accuracy. Although this introduces additional computational overhead, the enhanced model capacity enables superior segmentation performance, achieving a reasonable balance between accuracy and complexity.}

\begin{table}[!htbp]
\centering
\resizebox{0.7\textwidth}{!}{ 
\begin{tabular}{@{}lcccccccc@{}}
\toprule
\multirow{2}{*}{Complexity} & \multicolumn{7}{c}{Methods} \\
\cmidrule{2-8}
 & V-Net  & UA-MT & SASSNet & DTC & URPC & MC-Net+ & SASNet \\
\midrule
Para. (M) & 9.18  & 9.18 & 9.44 & 9.44 & 5.85 & 9.44 & 11.26 \\
FLOPs (G) & 46.85 &  46.85 & 46.88 & 46.88 & 69.36 & 46.88 & 67.31  \\
\bottomrule
\end{tabular}}
\caption{\textcolor{Revision}{Comparisons of computational complexity of inference between SASNet and other methods on the LA dataset.}}
\label{tab:complexity}
\end{table}

\section{\textcolor{Revision}{Discussion}}
\textcolor{Revision}{Although SASNet achieves competitive results, it is important to consider certain limitations. The current evaluation is restricted to three public datasets (LA, Pancreas-CT, and BraTS), which mainly cover cardiac, abdominal, and brain regions. This leaves open the question of whether the method can generalize to multi-organ segmentation tasks or more heterogeneous clinical cohorts. Moreover, SASNet has been primarily designed and validated for 3D volumetric data, and it has not been evaluated on 2D slice-based data or other imaging modalities (e.g., ultrasound, X-ray), so its applicability in these scenarios remains uncertain and warrants further investigation.}

\textcolor{Revision}{In light of these limitations, future work could extend SASNet to a broader range of segmentation tasks and systematically assess its generalization across different imaging modalities, scanner types, and multi-center clinical datasets. Moreover, integrating SASNet with other advanced techniques, such as graph-based embeddings or Transformer architectures, may further enhance its robustness and practical applicability.}

\section{Conclusion}
In this paper, we propose a novel semi-supervised segmentation network based on scale invariance (SASNet). The SASNet incorporates a new multi-scale learning\textemdash {scale-aware adaptive learning} into semi-supervised learning, which is combined with the SDM algorithm to achieve consistent learning of ensemble regression prediction and individual regression prediction. Additionally, we propose a view variance enhancement mechanism combined with multi-scale branches, which emulates annotation variations. This augmentation enhances the robustness of semi-supervised learning, thereby elevating the segmentation performance of the model. The evaluation of three public datasets including the left atrium (LA) dataset, the Pancreas-CT dataset, and the BraTS dataset shows that SASNet outperforms existing semi-supervised methods and achieves performance comparable to fully supervised methods. Future studies can investigate the potential of applying SASNet to various segmentation tasks and examine the feasibility of integrating SASNet with other techniques. 

 \bibliographystyle{elsarticle-num} 
 \bibliography{main}

\end{document}